# Gamification of Motor Imagery Brain-Computer Interface Training Protocols: a systematic review

Fred Atilla[1*], Marie Postma[1], Maryam Alimardani[2]

[1]*Department of Cognitive Science and Artificial Intelligence, Tilburg University, Tilburg, Netherlands*

[2]*Department of Computer Science, Vrije Universiteit Amsterdam, Amsterdam, Netherlands*

***Corresponding Author:**

Department of Cognitive Science and Artificial Intelligence, Tilburg University

Warandelaan 2, 5037AB

Tilburg, NETHERLANDS

e-mail: f.atilla@tilburguniversity.edu

**Author Note**

Fred Atilla https://orcid.org/0000-0003-1061-4420

Marie Postma https://orcid.org/0000-0001-8082-4883

Maryam Alimardani https://orcid.org/0000-0003-3077-7657

**Abstract**

Current Motor Imagery Brain-Computer Interfaces (MI-BCI) require a lengthy and monotonous training procedure to train both the system and the user. Considering many users struggle with effective control of MI-BCI systems, a more user-centered approach to training might help motivate users and facilitate learning, alleviating inefficiency of the BCI system. With the increase of BCI-controlled games, researchers have suggested using game principles for BCI training, as games are naturally centered on the player. This review identifies and evaluates the application of game design elements to MI-BCI training, a process known as gamification. Through a systematic literature search, we examined how MI-BCI training protocols have been gamified and how specific game elements impacted the training outcomes. We identified 86 studies that employed gamified MI-BCI protocols in the past decade. The prevalence and reported effects of individual game elements on user experience and performance were extracted and synthesized. Results reveal that MI-BCI training protocols are most often gamified by having users move an avatar in a virtual environment that provides visual feedback. Furthermore, in these virtual environments, users were provided with goals that guided their actions. Using gamification, the reviewed protocols allowed users to reach effective MI-BCI control, with studies reporting positive effects of four individual elements on user performance and experience, namely: feedback, avatars, assistance, and social interaction. Based on these elements, this review makes current and future recommendations for effective gamification, such as the use of virtual reality and adaptation of game difficulty to user skill level.

## 1. Introduction

The past decade has seen growing interest in the development of brain-computer interfaces (BCI). This type of interface allows users to interact with external devices and applications using solely signals generated from their brain activity, bypassing traditional motor inputs [1]. Initially, BCI were developed as a promising assistive technology for physically impaired users [2,3], helping them with essential activities such as communication [4], locomotion [5], and neurorehabilitation [6]. With the exponential increase of BCI studies [3,7], this emerging technology is also showing use in applications for able-bodied users [8], such as art [9], gaming [10], education [11], and marketing [12].

Regardless of the target audience or application, BCI are complex systems that require extensive training for effective use in real-world settings. This involves both training the user to generate clear and stable brain signals associated with mental tasks or states, and training the system (i.e., machine learning) to recognize the task/state from the user's signals [13]. However, this co-adaptation of the user and BCI system learning from each other still relies on suboptimal protocols [14,15], as a large portion (15-30%) of users remain incapable of controlling BCI effectively [16,17]. This issue, referred to as BCI inefficiency, is most prevalent for BCI that rely on motor imagery (MI-BCI) as they require active modulation of brain activity by the user [15,18].

To alleviate this issue, the BCI research community has proposed more user-centered approaches that rely on principles from human-computer interaction [19–24]. This is supported by the numerous user factors found to play a role in MI-BCI performance [25–30], whether it be a cognitive (e.g., spatial ability), psychological (e.g., motivation), or demographic factor (e.g., gender). One especially promising strategy that could engage users and promote learning of the MI task is the use of games or game-like training [13,20,31–33]. Making a non-game activity such as BCI training game-like can be achieved using game design elements such as rewards, avatars, and narratives [34]. This process called gamification has been shown to be beneficial to participants in other settings, such as education [35,36], healthcare [37], and corporate training [38].

While there has been an increase in BCI games in recent years [39], studies mainly focus on using the BCI as a controller for gameplay, rather than using the game for motor imagery training and better BCI control. To date, only one paper reviewed the potential of gamification for MI-BCI training [40]. While general gamification was found to positively impact MI-BCI performance, the review did not examine individual game design elements, making it difficult to determine how gamification was employed in past MI-BCI studies and how it can be employed effectively in future MI-BCI studies. In addition to user performance, it is also important to understand how a gamified strategy affects the user's experience, such as their motivation and engagement, as this could elucidate the mechanisms by which gamification impacts BCI performance. Therefore, there is still a need in the BCI field for a comprehensive systematic review of gamification approaches applied to MI-BCI training.

This review aims to identify and evaluate the application and effectiveness of gamification to MI-BCI training. With a focus on the design of protocols employed in previous MI-BCI studies, we aim to synthesize methodological guidelines, challenges, and potential research opportunities for the gamification of MI-BCI training protocols. Through a systematic literature review, the following research questions are answered:

1. *How is gamification applied to MI-BCI training?*
2. *Is there evidence supporting the effectiveness of gamification of MI-BCI training in terms of improving relevant user outcomes?*

The first research question involves examining *what* game elements were used in previous MI-BCI training protocols and *how* they were implemented. The second question involves identifying *whether* and *how* game elements influenced training outcomes related to the user's experience, performance, and MI-related brain activity. This is achieved through a systematic literature review, discussed more in Section 3. Before that, Section 2 presents background and related work on the concepts of BCI and gamification. Following the methodology, Section 4 summarizes our results, followed by effective gamification recommendations based on the findings in Section 5.

# 2. Background

## 2.1. Motor-Imagery Brain-Computer Interfaces

A brain-computer interface or brain-machine interface is a "new, non-muscular communication and control channel" between the brain and an external device such as a computer or robot [3,41–43]. To illustrate this concept, Fig. 1 shows the common stages of a closed-loop interaction with a BCI system. Firstly, the system captures brain signals from a user while some cognitive process or mental state occurs. For signal acquisition, typically electroencephalography (EEG) is employed due to its high temporal resolution, low costs, portability, and noninvasiveness [42]. These signals are preprocessed per the utilized neuroimaging technique (e.g., noise reduction and artifact removal), followed by extraction of discriminative features that help identify the intended mental state using machine learning algorithms. The classified output can then be mapped to meaningful commands in an external application, such as movement of physical robots or virtual avatars [44,45]. Finally, information about the commands can be fed back to the user through different sensory modalities to inform them of their success, subsequently forming a closed-loop control system in which the user and the machine learn from each other.

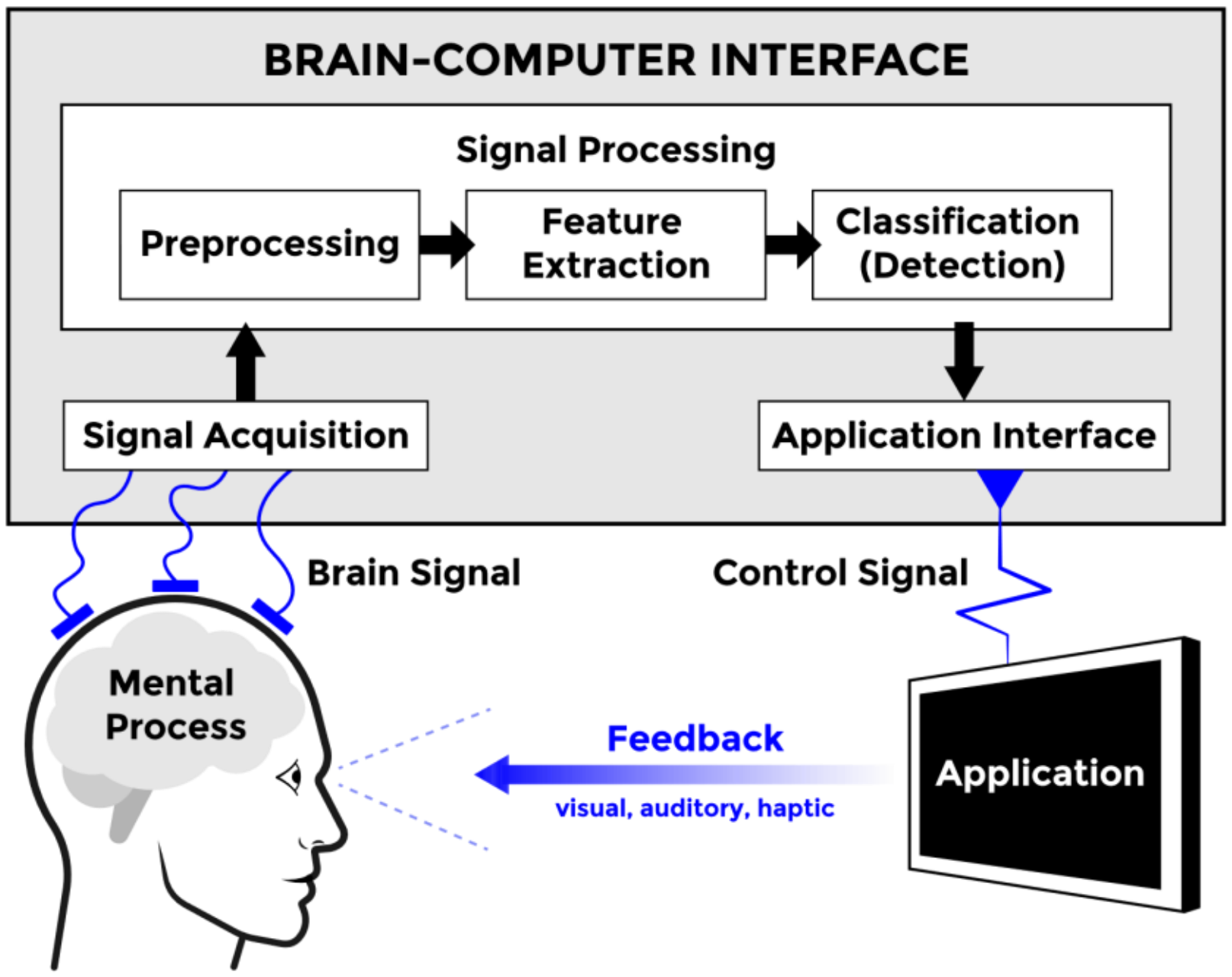

*Figure 1.* Overview of a closed-loop BCI system: acquired brain signals are preprocessed to extract discriminative features, which are classified by machine learning algorithms to detect user states or intentions for control of external applications. Finally, this information is fed back to the user.

The exact methods employed at each stage of the loop depend on the BCI paradigm. An established division by Zander et al. [46] categorizes BCI as either active, reactive, or passive. Active BCI are systems where the user consciously controls their brain activity by performing mental tasks without any external stimulation. For instance, users can imagine their hands grasping without actually executing any movements to induce similar brain activity as motor execution. Such motor imagery can then be discerned by the BCI system in real-time to make a prosthetic hand grasp. With reactive BCI, brain activity is elicited through external stimuli or events, allowing the user indirect control. A popular example is the P300 speller, which relies on a clear event-related potential that occurs ~300 ms after the user sees their desired letter light up on a grid of characters. Lastly, passive BCI analyze any ongoing brain activity that is not intentionally modulated and contains implicit information about the user's mental state, e.g., emotions or attention [47,48]. While each of these BCI require distinguishable brain activity from users, active BCI that rely on motor imagery (MI-BCI) need the most training as each user has unique brain activity that they must learn to modulate.

An MI-BCI works by detecting intended movements from neural oscillations connected to sensorimotor processes [42,49]. When users perform physical movements, oscillatory activity over sensorimotor areas called the sensorimotor rhythm (SMR) decreases in amplitude [50–52]. This amplitude change mostly occurs in the mu (8-13 Hz) and beta (14-30 Hz) bands, and is referred to as event-related desynchronization (ERD), which is followed by synchronization (ERS) after the movement is complete. This oscillatory activity not only occurs during execution but also during mental imagery of motor movements [53,54]. Therefore, users can consciously modulate their brain activity by imaging movements of body parts such as hands and feet. The resulting ERD/ERS patterns can be discriminated by machine learning algorithms to reliably detect what movement the user intended. For example, simple Linear Discriminant Analysis (LDA) can classify left-versus right-hand movements with 95% accuracy [55,56]. Obtaining consistently good accuracies, however, is challenging as the number of classes and dimensionality of the data increase. Thus, there is still a substantial focus on finding better signal processing and classification algorithms (for a review, see [41,57,58]).

To train the classification algorithm, labeled brain activity associated with motor movements is required. At the same time, the user needs to be trained to produce clear and stable patterns of brain activity for different motor movements through mental imagery. A standard training paradigm for this interdependent process is the Graz protocol [59]. Fig. 2 shows the temporal structure of one trial in the Graz protocol, which takes 8 seconds. After 3 seconds of fixation, a horizontal arrow pointing to the left or right cues the user which movement to imagine, either left or right hand, respectively. The user performs MI for 4 to 5 seconds while their sensorimotor activity is captured through a neuroimaging technique such as EEG. In this manner, clearly labeled signal data is gathered for the BCI system to train on. Once the classifier is trained, it can make decisions and provide feedback to the user in the form of a horizontal bar extending to the left or right side, whose length reflects the classification strength. No such feedback is provided in the initial calibration session where data for the training of the classifier is collected.

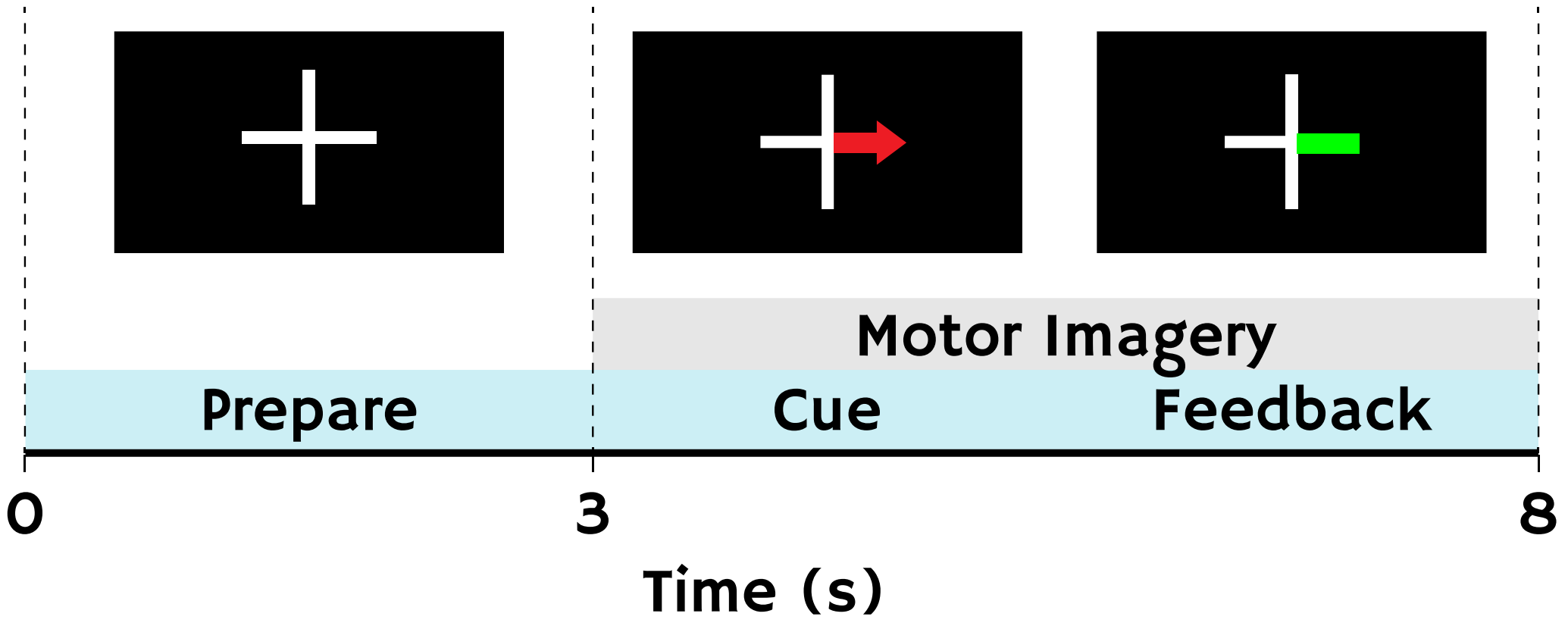

*Figure 2.* Experimental trial used in the standard Graz protocol for training of MI-BCI control.

Despite such user training, a growing body of literature reports that 15-30% of users are incapable of effective BCI control beyond chance-level accuracies or 70% accuracy for binary classification [15,17,60]. This issue which was initially termed 'BCI illiteracy' is especially prevalent among MI-BCI users [15,18]. However, to more accurately reflect the challenges faced by the users of BCI systems, recent studies use the terms 'BCI deficiency' or 'inefficiency' [15,17,60]. This shift in terminology highlights the flawed assumption that some users are inherently incapable of BCI control; rather, they struggle to learn BCI control due to suboptimal training protocols, unclear instructions, and repetitive trials [14,17]. It is crucial to ensure users are motivated and attentive during training, as many user factors such as mental states and other psychological factors can affect BCI performance [13,27,29,61,62]. Here, a more user-centered approach through gamified training might be beneficial.

**2.2. Gamification**

The user-centered approach is central to game design, where choices are made by taking the user's (i.e., player) subjective experience and attitude into account, for example through playtesting [63]. Elements from game design can also be used in non-game activities, a process known as gamification [34,64]. Unlike full-fledged games for the purpose of entertainment, a non-game activity, such as learning a new subject, is gamified by incorporating game elements, such as points for each completed exercise. The goal of gamification is to facilitate learning and motivate participants through genuine fun and exploration over mere obligatory participation, e.g., for study credits [34,65]. Its effectiveness has been extensively explained through psychological theory [34,65–69]: in general, game elements impart motivation (i.e., motivational affordances), which influences psychological outcomes (e.g., engagement, pleasure), which in turn influences behavioral outcomes (e.g., learning, participation).

While the goal of gamification is clear, there is still a lack of consensus regarding its operationalization into game elements [70,71]. Nevertheless, the many gamification taxonomies put forth over the past decade share at least some elements. To be as inclusive as possible and account for different views of gamification, this review incorporates 14 commonly employed game elements (Table 1). Gamification can for instance be implemented by providing users with clear goals to guide them, challenges to overcome, rewards to reinforce successful behaviors, or information on their performance [72,73]. This process must be carried out with careful attention, as gamification requires more than superficially adding elements such as points, badges, and leaderboards to a rudimentary task [74].

For meaningful gamification, many frameworks highlight the importance of first identifying useful game elements and evaluating their effects in an applied system [64,75–77]. Such research has been performed for many activities, shedding light on the positive impacts of gamification in education [35,36], healthcare [37], physical exercise [78], tourism [79], management [80], therapy [81,82], rehabilitation [83], and corporate training [38]. For example, in education, the impact of gamification has most often been studied in relation to user motivation, engagement, and academic performance, with numerous studies reporting positive effects on these outcomes (for a review, see [35]). Furthermore, points, badges, leaderboards, and narratives were the most frequently applied game elements to educational activities. What elements are most frequently applied and how they affect learning differs per field [37,67,78,81,84], suggesting the need for dedicated research on gamification for the task of MI-BCI training.

Table 1

*Common game elements found in existing game design and gamification taxonomies.*

| Element | Description | Based on |
|---|---|---|
| Feedback | Immediate visual, auditory, or haptic information users receive about their (inter)actions within the (virtual) environment in real-time. Feedback informs users whether their actions were successful, allowing them to learn from previous actions and modify their behavior. | [69,72,84–88] |
| Goals | The reasons for which a user interacts with the environment and ultimately the purpose of their actions. Also known as objectives. Multiple goals can be set at different time scales. Related to goals are achievements, which are optional tasks that can be completed through a specific set of actions, e.g., completing a certain number of trials. | [69,72,85,89] |
| Avatar | Visual representation of the user within the (virtual) environment. This can for instance be a virtual avatar, anthropomorphized animal character, or physical robot. | [66,85,88,90,91] |
| Challenges | Activities or constraints that provide a cognitive or behavioral challenge to the user and must be overcome to achieve goals. | [69,72,88,89] |
| Rewards | Incentives given to a user within the environment for performing certain actions successfully. These incentives provide positive reinforcement. Common rewards are points (a numeric unit that quantifies user performance) and badges (visual icons representing achievements). | [66,68,69,73,87,88, 90,91] |
| Progress | Visual indication of how far the user has advanced through the environment and where they stand in relation to achieving their goals. | [72,88,89,91] |
| Score | Visual indicator of the user’s performance on a success criterion, e.g., number of received points. | [72,90] |
| Timer | Visual indicator of time taken or left to complete tasks. | [85,89,92] |
| Assistance | Any form of aid provided by the system or other people to help the user progress. Common forms include tutorials that instruct how to use the system and adaptation that adjusts the task difficulty to the user’s skill. | [72,86,88,93] |
| Levels | Hierarchical layers that the user can advance through, commonly in the form of new locations, new levels of difficulty, or both simultaneously. | [73,85,89,91] |
| Social Interaction | The user interacts with other users, either in the form of collaborating (working together) or competing (facing each other) to achieve a shared/common goal, e.g., solving a puzzle. | [88–90,92,93] |
| Leaderboard | Display of a user’s rank compared to other users based on a particular task/criterion, e.g., points or completion time. | [66,73,84,85,90,91] |
| Personalization | Allowing the user to transform elements of their experience, such as their avatar, according to their preferences. | [85,87] |
| Narratives | Events linked together within the environment to give context and meaning to user actions (beyond the mere pursuit of points and achievements). Also known as storylines or game fiction. | [72,85,88,90,94] |

### 2.3. BCI and Gamification

The convergence of BCI and games has been reviewed considerably. Most past work focuses on the application of BCI to games, with two general approaches to improving game interaction [32,95–100]. Firstly, passive BCI can be used to monitor the mental states of users. This information can then be used in various ways to adjust aspects of the game, such as lowering the difficulty when users lose engagement or become frustrated. The second way is to use (re)active BCI as a more natural, hands-free method of control. The latter has been found to be most important to users, game developers, and researchers according to a survey by Ahn et al. [10]. This is confirmed by several reviews that found the MI paradigm to be the most employed paradigm in BCI games [95,98].

In games, the MI-BCI is typically used to control the movement of a virtual object or character inside a 2D or 3D environment [95,98,101]. This is intuitive as the imagined movements are analogous to in-game movements [39]. Categorizing BCI games by genre, MI-BCI have also been found to be the most frequent paradigm for action games, which in turn are the most popular genre for BCI games [39,98]. However, this genre involves fast-paced gameplay that requires quick reaction times, therefore compromises have to be made for BCI adaptation. For instance, the pace needs to be reduced with designated time windows for commands (i.e., cue-paced) and the number of possible actions decreased to allow accurate control while keeping gameplay enjoyable [98]. To integrate BCI better into gaming, Cattan et al. [102] made several recommendations: provide users with guidance, set goals, make multiple users collaborate, and incorporate stimuli into the game environment. While these recommendations were made for P300-based BCI, they could extend to other BCI paradigms. Similarly, Ferreira et al. [95] provided a design rationale for decisions made when designing BCI games, e.g., headset and paradigm choice, with negatives and positives to consider for each decision. This is especially useful as most studies develop their own BCI games rather than adapting existing games [10,39].

Contrary to the use of BCI for games, the application of games to general BCI training has been overlooked, with some suggestions made in passing by past reviews. Plass-Oude Bos et al. [32] proposed that boredom during calibration can be reduced by letting users train in a game environment and using tutorials or mini-games to provide clear instructions. Similarly, Vasiljevic and de Miranda [39] suggested gamifying the training stage through mini-games and adapting the game's difficulty as the user learns to play. They also advocated for more research into the effects of game elements, such as graphics and sound effects, on the user's control signal. Another suggestion for gamification was made in Marshall [98]: users can be assisted toward in-game objectives with guidance from objects. Finally, several works have suggested the potential of virtual reality (VR) to make BCI training more engaging and game-like: VR as a platform offers immersive and controlled environments with enriched feedback that can facilitate motor learning and imagery [96,101,103–106].

To the best of our knowledge, only two previous reviews focused on improving BCI training through game principles. Scherer et al. [107] proposed to replace standard training protocols with games to learn the "BCI skill". Games can increase user motivation and engagement, and ultimately BCI control as these user states in turn affect modulation of brain activity. Standard protocols were criticized as being too simplistic, not allowing users to explore different mental tasks or commands; presenting minimalistic graphics; providing obscure feedback in the form of numbers or abstract brain signals; providing no (sham) feedback during calibration; and burdening users with guessing their performance instead of quantifying it for them in real-time. Rather than adapting off-the-shelf games, the authors suggested either building a BCI skill training game (a serious game) or using game design elements in existing BCI protocols (gamification); the former option being more expensive and challenging. Other helpful game principles were to balance the user's skill and the game's difficulty level to incite a flow state, and to provide an environment with clear goals and self-explored challenges. Finally, a few BCI games were summarized, identifying the use of game elements such as goals and scores, but further investigation into gamification and its effectiveness was not performed.

Another review by Gao et al. [40] conducted an investigation into gamification for MI-BCI learning through a systematic review of 28 studies. Gamification was found to be viable for MI-BCI training with an average accuracy of 74% across studies. This finding is, however, of uncertain validity due to the averaging of accuracies from studies with distinct methodologies. Most studies reported performance outcomes, such as classification accuracy or game score, with few reports of psychological or cognitive outcomes. In terms of gamification, the 28 games were categorized into main genres and subgenres based on (1) perspective (third- or first-person), (2) visual type (2D, 3D, or VR environment), and (3) game content (task and goals to complete). In line with the aforementioned reviews, most studies used action games, with common subgenres being 2D games where users avoid obstacles and VR first-person games. None of the included studies, however, compared gamified training with a control condition. Therefore, it is still unclear whether gamifying standard MI-BCI training is beneficial, despite the ‘positive reports’ from the reviewed studies in Gao et al. [40]. In addition, common game design elements used in gamification (Table 1) were not identified or evaluated. This is necessary to establish which specific design elements positively contribute to (which) user outcomes. Thus, a systematic review identifying game elements and evaluating their effectiveness with empirical control is warranted.

## 3. Methods

To answer the research questions, a systematic literature review was conducted according to the Preferred Reporting Items for Systematic Literature Reviews and Meta-Analyses (PRISMA) protocol [108]. The PRISMA protocol helps streamline the identification, selection, and synthesis of relevant studies connected to the research questions. Henceforth, each step is described in further detail.

### 3.1. Search Strategy for Data Collection

To identify relevant publications, a systematic search was performed across several bibliographic databases. To get maximum coverage of the literature where the terminology was not always consistent, we conducted a search using a broad search query consisting of three sets of keywords related to BCI, the MI paradigm, and gamification, respectively (Table 2). The three groups were connected using Boolean logic with AND operators and the related terms within each group were connected using OR operators. The resulting search query was for example: (brain-computer interface OR brain-machine interface OR …) AND (motor imagery OR …) AND (game OR …). For each term in Table 2, variations were also included, e.g., ‘gamified’ for gamification. The full search queries and list of terms are available for reproducibility in the Supplementary Materials.

Because brain-computer interfaces intersect a range of disciplines, the search was performed on six popular bibliographic databases for the fields of medicine, social sciences, computer science, and engineering: Scopus, Web of science, PubMed (OVID), ProQuest, IEEExplore, PsycINFO (EBSCOhost). In each database, the terms were searched for in the abstract, title, and keywords. Searches were performed on 3 January 2023, resulting in a total of 578 publications found (Fig. 3) when filtering with the first two inclusion criteria, discussed next.

Table 2

*Terms used in the systematic search for gamified MI-BCI literature.*

| **BCI** | **&** | **PARADIGM** | **&** | **GAME** |
|---|---|---|---|---|
| brain-computer interface | | motor imagery | | game |
| brain-machine interface | | imagined movement | | videogame |
| human-computer interface | | kinesthetic imagery | | gamification |
| human-machine interface | | event-related (de)synchronization | | virtual reality |
| brain-controlled | | motor intention | | virtual environment |
| mind-controlled | | | | |
| neurofeedback | | | | |

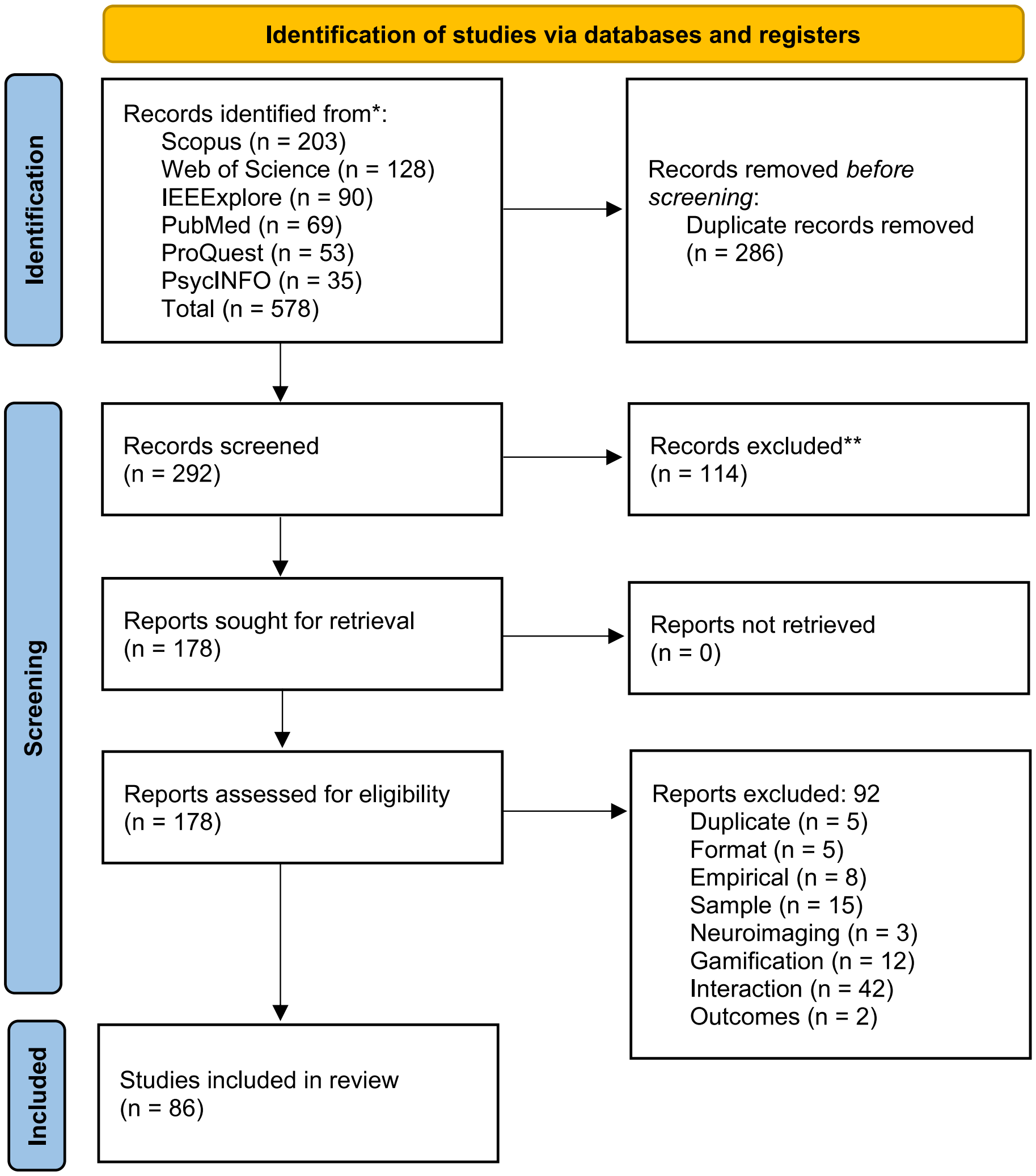

*Figure 3.* PRISMA flow diagram of search and screening of gamified MI-BCI training protocols.

### 3.2. Screening and Selection for Inclusion

The initial results were imported to Zotero (v.6.0.23), a reference manager. Duplicates were automatically removed using Zotero, with a few more duplicates manually removed during later screening, resulting in a total of 286 duplicates removed from data analysis. The 292 unique publications were screened by reading the title and abstract first to quickly identify relevant literature. Strict criteria were set to limit literature to the scope of this review and ensure the quality of included papers. Specifically, papers were included when they met all of the following criteria:

1. Format: English peer-reviewed journal articles, conference papers, or book chapters. Extended abstracts, talks, dissertations, and similar works were not considered.
2. Year: Published in the past decade (2012-2022) to include up-to-date BCI and gamification practices.
3. Empirical: Performed an empirical study with participants. Any methodology sufficed, whether qualitative, quantitative, or mixed. Sole proposals for a BCI system were thus excluded.
4. Sample: A minimum of 2 participants to ensure outcomes were not unique to one individual; single case studies, common in medical papers, were thus excluded. Participants could be of any demographic, e.g., healthy, stroke patients.
5. Neuroimaging: Brain signals were acquired through EEG or fNIRS, as these are portable, non-invasive, and allow the MI-BCI to be used in real time. Nevertheless, after screening, all included papers used EEG as the main method for signal acquisition.
6. Gamification: Participants performed a (serious) game, game-based, or gamified task, where either the authors explicitly stated such a task was used, or at least one game element was apparent from the task description or figures. Papers were excluded if their protocol was not described sufficiently to evaluate whether or how the task was gamified.
7. Interaction: Employed a motor imagery BCI paradigm where the participants used imagined movements to actively control an aspect of the gamified protocol. Such real-time interaction was accompanied by feedback and validated use of the system for active control. The study was excluded if there was simultaneous motor execution during MI, as that might confound any potential effects of gamification on solely MI. Studies with hybrid paradigms where motor execution was not involved were however included, e.g., MI and P300 signals.

8. Outcomes: Reported qualitative or quantitative results about the user's performance (e.g., BCI online accuracy), MI-related brain activity (e.g., mu rhythm, ERD), or experience (e.g., engagement, motivation, embodiment).

The first author screened all titles and abstracts based on these criteria, and the second and third authors screened a subset of 36 (20%) abstracts as a reliability check. Disagreements were resolved through discussion. Ultimately, this resulted in excluding 114 papers, keeping 178 for full-text screening. All papers were retrieved to assess their full text for eligibility. The same criteria above were used during the full screening, resulting in 86 final papers being eligible for analysis in this systematic review (Fig. 3). Most of these papers were journal articles and published in recent years (Fig. 4).

### 3.3. Information Extraction

The 86 eligible papers were coded and analyzed with respect to relevant aspects of their methodology and findings. Extraction categories were developed by considering systematic reviews on BCI and gamification. Information was extracted per paper and entered into a spreadsheet along the following dimensions:

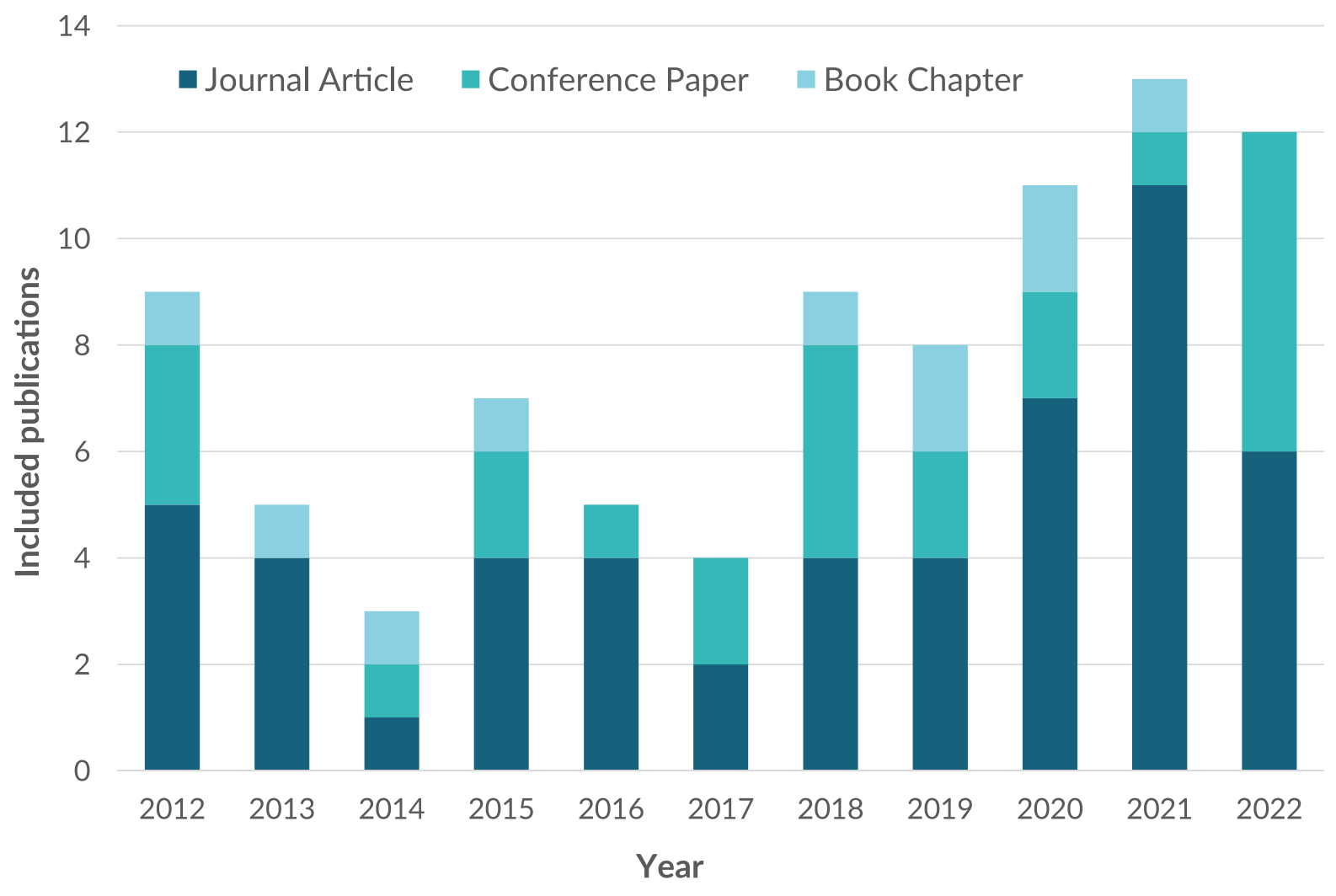

*Figure 4.* Number of papers included in this review per year and publication type.

1. Study design: sample size and characteristics, conditions, cue indicating target MI task, and BCI application.
2. Game design elements: the implementation of common game elements (Table 1). Where elements were not explicitly stated by the authors, the description and figures were used to evaluate whether an element was present in the protocol. For studies with multiple protocols, elements pertaining to the most gamified protocol are reported.
3. Main outcomes: to examine the effectiveness of gamification for MI-BCI training, the necessary outcomes and their measures were identified. Outcomes could fall under one of three categories: (1) user performance, e.g., classification accuracy, game score; (2) user experience, e.g., mental states, motivation, usability; and (3) users' MI ability, either measured through a questionnaire or neural activity.

Information extraction was performed by the first author and iteratively improved upon through discussions with the second and third author. Missing, unclear, and irrelevant information were noted as not available/applicable (N/A). Information from the different categories was narratively synthesized by summarizing the implementation and reported effectiveness (if empirically studied) of each game element on the main outcomes. In doing so, information was summarized per paper, but it should be noted that some studies employed the same protocol or a slightly modified version (typically studies by the same researcher or lab). To concisely answer the research questions, the results that follow convey extracted information regarding the latter two categories (game design elements and main outcomes). For extracted information regarding the study design, we refer readers to the Supplementary Materials (Supplementary Table 3).

# 4. Results

## 4.1. Gamification in General

Table 3 presents all 86 studies included in this review together with a summary of the game elements that were implemented. Among the 14 common game elements that were used for data extraction, MI-BCI training protocols were most often gamified using feedback, goals, and avatars (Fig. 5). The former is not out of the ordinary as real-time interaction, one of our inclusion criteria, is inherently accompanied by feedback in BCI training. The amount of gamification varied considerably, with some studies implementing only two elements and others nine. For instance,

Ortner et al. [109] used gamified *feedback* of a virtual *avatar* moving its arms in accordance with the user's motor imagery (2 elements; Fig. 6a). On the other end, in Castro-Cros et al. [110] users received real-time *feedback* of a virtual *avatar* that moved its arms to block a mouse from taking a piece of cheese, giving users a *goal* embedded in a *narrative*, and visually indicating their *progress*, *score*, and game *time* (7 elements; Fig. 6b). This study was also one of the few that compared their gamified protocol to a control protocol. It was found that gamification had no effect on MI-BCI performance, but users did rate the gamified experience positively, being highly satisfied, and stating it improved concentration and alleviated boredom.

Nevertheless, other results on gamification of MI-BCI training are mixed and hard to interpret due to a lack of empirical control. Li et al. [111] compared a gamified protocol in an immersive 3D Tetris environment (Fig. 6c) to an analogous 2D game (Fig. 6d) and found that the former improved user performance. Likewise, in Hougaard et al. [112] users had more consistent control and preferred a gamified kiwi runner task with a narrative (Fig. 6f)) to a simple ball squeezing task (Fig. 6e). Many of the remaining studies included in this review only had one condition, namely the one where a gamified protocol was used. Users in these studies often reached effective control, that is, above chance or the commonly used threshold of 70% for binary classification, e.g., left- versus right-hand imagery [45,113–126]. Furthermore, users often rated their experience with the gamified training positively [45,117,127,128], e.g., enjoyable [115,129], immersive [130], engaging [131], and good usability [132]. However, not all findings are positive as some users found their training boring and not motivating [133], high on workload [132], or difficult to control the BCI [115].

Overall, gamification seems to positively influence user performance and experience, but it should be noted that without control conditions many of these findings do not constitute strong evidence for its general efficacy. Furthermore, even with such control conditions — as in the case of Castro-Cros et al. [110], Li et al. [111], and Hougaard et al. [112] — it is difficult to determine the effect of individual elements due to the simultaneous presence of multiple game elements. Comparison of experimental conditions with and without specific game elements is crucial to making recommendations for effective gamification of MI-BCI training. Henceforth, the results discuss each element in detail; specifically, how often they were implemented, how they were implemented, how often their effectiveness was studied, and their reported effects on user outcomes (summarized in Figure 7).

Table 3

*List of all 86 MI-BCI studies included in this review with a summary of the implemented game elements. Additional extracted information regarding the study design is available in the Supplementary Materials. Feedback modality: V = Visual, H = Haptic, A = Auditory. Feedback platform: HMD = head-mounted display, M = monitor, RW = real world. Goals: short-term (ST), long-term (LT). Avatar: 1 = first-person, 3 = third-person, H = human, N = non-human, X = no appearance. Progress: B = bar, N = numeric, I = in-game. Rewards: P = points, U = unique. Timer: U = counts up, D = counts down.*

| | | Feedback | | | | | | | | | | | | | | | |
|---|---|---|---|---|---|---|---|---|---|---|---|---|---|---|---|---|---|
| **Ref** | **Study** | **Modality** | **Platform** | **Content** | **Goals** | **Avatar** | **Challenges** | **Progress** | **Rewards** | **Score** | **Timer** | **Assistance** | **Levels** | **Social Inter.** | **Leaderboard** | **Personalization** | **Narrative** |
| *[113]* | Achanccaray et al. (2018) | V | HMD | virtual arm extension | n/a | 1H | | | | | | | | | | | |
| *[134]* | Achanccaray et al. (2019) | V | M, HMD | virtual arm extension | n/a | 1H | | | | | | | | | | | |
| *[135]* | Achanccaray et al. (2021) | V, H | HMD | virtual arm moves + real forearm electro-tactile stimulation | grab ball (ST) | 1H | | | | | | | | | | | |
| *[136]* | Alchalabi & Faubert (2019) | V | VR Icube | tunnel movement | n/a | | | | | | | | | | | | |
| *[137]* | Alchalabi et al. (2021) | V | HMD | avatar moves | reach target arrow (ST) | 1H | | | | | | | | | | | |
| [44] | Alimardani et al. (2015) | V | RW | humanoid robot hand grasps ball | grasp illuminated ball (ST) | 1H | | | | | | | | | | | |
| *[114]* | Arpaia et al. (2021) | V, H | M | virtual ball moves + haptic suit vibrates | move ball to line (ST) | | | N | | | | | | | | | |
| *[138]* | Arpaia, Coyle, et al. (2022) | V, H | M | virtual ball moves + haptic suit vibrates | move ball to line (ST) | | | N | | | | | | | | | |
| *[139]* | Arpaia, D'Angelo, et al. (2022) | V, A | HMD | virtual arm lifts platform + ball rolls + sound effect if goal | move ball to opponent's goal (ST) to score as many points as possible (LT) | 1H+ 3H | ✓ | | P | ✓ | | ✓ | | | | | |
| *[140]* | Asensio-Cubero et al. (2016) | V | M | character moves + colored arrow | collect coin and avoid being bitten by snakes (ST) to collect as many coins as possible (LT) | 3H | ✓ | N | P | ✓ | | | | | | | |
| *[115]* | Badia et al. (2013) | V | HMD | virtual arm hits sphere | hit incoming spheres (ST) | 1H | | | P | ✓ | | ✓ | | | | | |
| *[141]* | Battison et al. (2015) | V | M | cursor moves | move bar to target position (ST) | | | | | | | ✓ | | | | | |
| *[142]* | Bigirimana et al. (2020) | V | M | neuron moves | catch incoming spike (ST) | 3N | ✓ | | P | | | | | | | | |
| *[143]* | Blanco-Mora et al. (2022) | V | M, HMD | virtual hand rows | row a boat (ST) to indicated flags to collect as many flags as possible (LT) | 1H | ✓ | | P | ✓ | D | ✓ | | | | | |
| *[144]* | Bois et al. (2021) | V | M | neuron moves | catch incoming spike (ST) | 3N | ✓ | | P | | | | | | | | |
| *[145]* | Bonnet et al. (2013) | V | M | bar + ball moves | score a goal (ST) to win against opponent or score as many as possible together (LT) | | ✓ | B | P | | | | | ✓ | | | |
| *[116]* | Bordoloi et al. (2012) | V | M | yellow square moves + command as text | reach maze exit (LT) | | ✓ | I | | | | | | | | | |
| *[146]* | Cantillo-Negrete et al. (2019) | V, H | M, RW | virtual hand grasps or forced user hand grasp | n/a | | | | | | | | | | | | |
| *[147]* | Cardoso et al. (2021) | H | M, RW | passive pedaling | n/a | | | | | | | | | | | | |
| *[110]* | Castro-Cros et al. (2020) | V, H | M | virtual arm moves + FES | block mouse from grabbing cheese (ST) to preserve as much cheese as possible (LT) | 1H | | I | | ✓ | ✓ | | | | | | ✓ |
| *[148]* | Choi et al. (2020) | V | HMD | move through track + virtual hand grasps | reach end of track (LT) | 1H | | | | | | | | | | | |
| *[117]* | Cunha et al. (2021) | V | M | blue bar and stars fill up, hammer size increases and hits ghost | hit ghost (ST) to collect as many stars as possible (LT) | | ✓ | | P | ✓ | | | | | | | |
| *[149]* | Da Silva-Sauer et al. (2015) | V, A | Projector + Polarized glasses | smiley face + bell or error sign + buzzer | go into indicated direction (ST) to navigate through park (LT) | 1X | | | | | | | | | | | |
| *[150]* | Davis et al. (2022) | V | HMD | virtual arm grasp | n/a | 1N+ 3N | | | | | | | | | | | |
| *[151]* | Duan et al. (2015) | V | 2x M | virtual or real robot grasp | grasp object (ST) | 3N | | | | | | | | | | | |
| *[152]* | Espinosa et al. (2012) | V | VR PowerWall | virtual hand open/close | n/a | 1H | | | | | | | | | | | |
| *[153]* | Ferrero et al. (2021) | V | HMD, M | movement through virtual corridor | n/a | 1X | | | | | | | | | | | |
| *[154]* | Hasan & Gan (2012) | V | M | classifier output + letters + stick figure | guess correct letters (ST) to form the secret word to prevent the hangman (LT) | | ✓ | I | | | | | | | | | |
| *[112]* | Hougaard et al. (2022) | V | M | kiwi character jumps over or hits grass | jump over obstacle (ST) to reach nest to protect eggs from predator (LT) | 3N | ✓ | | | | | | | | | | ✓ |
| *[118]* | Huang et al. (2012) | V | M | wheelchair icon moves | reach target side (ST) | 3N | ✓ | | | | | | | | | | |
| *[155]* | Jochumsen et al. (2022) | V | M | character reels in or misses fish | reel in fish (ST) to catch as many as possible (LT) | 3H | ✓ | B | U | | | ✓ | | | | | |
| *[156]* | Juliano et al. (2020) | V, A | M, HMD | virtual arm moves + auditory tone | bring arm to position to hit beach ball (ST) | 1H | | | | | | | | | | | |
| *[157]* | Kapitonova et al. (2022) | V | HMD | virtual robot animation | match robot color to catch it (ST) to collect as many points as possible (LT) | 1H | | | P | | | | ✓ | | ✓ | | |
| *[130]* | Karácsony et al. (2019) | V | HMD | object explodes due to hand grasp or foot kick | catch falling fruit or kick ball (ST) | 1H | | B | | ✓ | | | | | | | |
| *[119]* | King et al. (2013) | V | M | avatar walks | move to next NPC (ST) to complete all stops (LT) | 3H | ✓ | N | P | | | | | | | | |
| *[158]* | Kjeldsen et al. (2021) | V, A | HMD | virtual hand pops balloon or buzzer sound | grasp balloon to pop it (ST) | 1H | | | | | | | | | | | |
| *[159]* | Kosmyna et al. (2014) | V | RW | drone moves + classification output | move drone to target position (LT) | 3N | | B | | | | | | | | | |
| *[160]* | Kosmyna et al. (2015) | V, A | M | rabbit or wolf animations + sound | save rabbit or shoot wolf (ST) to reach home (LT) | 1X | ✓ | N | | | | | | | | | ✓ |
| *[120]* | Kreilinger et al. (2016) | V, A | M | car moves + car size and color change + beeping sounds | hit coin and avoid barrier (ST) to collect as many possible (LT) | 3N | ✓ | N | P | ✓ | | | | | | | |
| *[161]* | Lechner et al. (2014) | V | VR PowerWall | virtual hand open/close | n/a | 1H | | | | | | | | | | | |
| *[162]* | Leeb et al. (2013) | V | M, CAVE | penguin jumps | collect fish (ST) to get as many points as possible (LT) | 3N | ✓ | | P | | | | ✓ | | | | |
| *[111]* | Li et al. (2017) | V | M | Tetris block moves | move block to prevent it from stacking and place it to form lines (ST) to stay | | ✓ | N | P | ✓ | | | ✓ | | | | |
| | | | | | alive as long as possible and maximize score (LT) | | | | | | | | | | | | |
| *[131]* | Liu et al. (2020) | V, A | M, RW | drone moves + virtual water jar fills, fireworks + celebration sounds at goal | fly to target height (ST) | 3N | ✓ | | | | | ✓ | | | | | |
| *[163]* | Long, Li, Wang, Yu, Pan, et al. (2012) | V, H | M, RW | virtual (2D) or real wheelchair moves | move car to target (ST) to hit all targets in order (LT) | 3N | ✓ | N | | | | | | | | | |
| *[164]* | Long, Li, Wang, Yu, & Pan (2012) | V | M | virtual (2D) wheelchair moves | move car to target (ST) to hit all targets in order (LT) | 3N | ✓ | N | | | | | | | | | |
| *[165]* | Mahmood et al. (2021) | V | M, HMD | text + cube disappears + EEG activity | score points (LT) | | | | P | ✓ | D | | | | | | |
| *[166]* | McCreadie et al. (2014) | V, A | M, RW | basket or spaceship moves, or asteroid and spaceship sounds | move spaceship to avoid incoming asteroid (ST) | 3N | ✓ | | P | ✓ | U | | ✓ | | | | |
| *[121]* | McMahon & Schukat (2018) | V | HMD | 3D shape rotation | rotate puzzle to correct position (LT) | | ✓ | | | | | | | | | | |
| *[167]* | Miao et al. (2020) | V, H | M | virtual arm moves + FES muscle contraction | n/a | 1H | | | | | | | | | | | |
| *[168]* | Mladenović et al. (2021) | V, A | M | penguin moves and hits fish + accompanying sounds | collect fish (ST) to score as many points as possible (LT) | 3N | ✓ | B | P | ✓ | U | ✓ | | | | | |
| *[169]* | Mladenović et al. (2017) | V, A | M | penguin moves and hits fish / squid + sounds | collect fish/squid (ST) to score as many points as possible (LT) | 3N | ✓ | B | P | ✓ | U | | | | | | |
| *[127]* | Morone et al. (2015) | V | M | virtual hand moves | n/a | 1H | | | | | | | | | | | |
| *[170]* | Myrden & Chau (2015) | V | M | 2D character moves | navigate to maze exit (LT) | 3H | ✓ | N | | | | | ✓ | | | | |
| *[171]* | Nicolelis et al. (2022) | V, H | HMD | avatar moves + bar + forearm vibration | move to target position (ST) | 1H | | | | | | | | | | | |
| *[172]* | Nierula et al. (2021) | V, A | HMD | virtual arm moves to button or error sound | press button (ST) | 1H | | | | | | | | | | | |
| *[173]* | Novak et al. (2018) | V | M | robot avatar moves | choose correct action to overcome obstacles and speed up (ST) to finish race fastest (LT) | 3N | ✓ | B | U | | U | | | ✓ | ✓ | | |
| *[174]* | Ortner et al. (2012) | V | VR PowerWall | virtual hand open/close | n/a | 1H | | | | | | | | | | | |
| *[109]* | Ortner et al. (2013) | V, H | RW, VR PowerWall | 1: robot closes/opens user hand<br>2: virtual hand open/close | n/a | 1H | | | | | | | | | | | |
| *[175]* | Perdikis et al. (2018) | V | M | robot avatar moves | choose correct action to overcome obstacles and speed up (ST) to finish race fastest (LT) | 3N | ✓ | B | U | | U | | | ✓ | ✓ | | |
| *[176]* | Petrushin et al. (2018) | V, H | M | mouse button click + pin pushes against finger | move to next target position (ST) to finish whole track (LT) | 1N | | N | | | | | | | | | |
| *[122]* | Ponferrada et al. (2018) | V | M | robot avatar movements + classifier output | choose correct action to overcome obstacles and speed up (ST) to finish race fastest (LT) | 3N | ✓ | B | U | | | | | | | | |
| *[177]* | Pradhapan et al. (2018) | V | RW | toy car moves | steer car (ST) to complete laps on racing track (LT) | 3N | ✓ | | | | | | | ✓ | | | |
| *[123]* | Prapas et al. (2022) | V | M | avatar moves on lane | collect coins (ST) | 3N | | | P | ✓ | | | | | | | |
| *[178]* | Ren et al. (2020) | V | M | virtual leg kicks ball | kick ball (ST) | 1H | | | | | | | | | | | |
| *[124]* | Saeedi et al. (2016) | V | M | platform moves and turns green or red | move platform under falling NPC to save them (ST) | | | | | ✓ | | ✓ | | | | | |
| *[179]* | Sakamaki et al. (2022) | V, H | M | 1: bar vs robot moves user's hand<br>2: character moves vs robot moves user's hand | n/a | 3N | | | | | | | | | | | |
| *[180]* | Sebastián-Romagosa et al. (2020) | V, H | M | virtual hand movement + wrist electro-stimulation | n/a | 1H | | | | | | | | | | | |
| *[181]* | Shane et al. (2022) | V | HMD | virtual arm moves | move to target (ST) | 1H | | | | | | | | | | | |
| *[182]* | Škola & Liarokapis (2018) | V | M, HMD | virtual arm pushes button | press button (ST) | 1H | | N | | | | | | | | | |
| [33] | Škola et al. (2019) | V, H | HMD | virtual hand pull lever shooting asteroid + user hand vibrates | pull lever to shoot asteroid (ST) to protect Earth (LT) | 1H | | N | | ✓ | | ✓ | | | | | ✓ |
| *[183]* | Tabernig et al. (2021) | V, H | RW, M | avatar foot dorsiflexion or real foot dorsiflexion with FES | n/a | 3H | | | | | | | | | | | |
| *[133]* | Taherian et al. (2017) | V | M | virtual cube moves | get puzzle piece (ST) complete puzzle (LT) | | | I | U | | | | | | | | |
| *[128]* | Tezza et al. (2019) | V | M | virtual drone moves | finish race as fast as possible (LT) | 3N | | | | | U | ✓ | | | ✓ | ✓ | |
| *[129]* | Tezza, Caprio, Garcia, et al. (2020) | V | M | virtual drone moves | finish race as fast as possible (LT) | 3N | | | | | U | | | | | ✓ | |
| *[184]* | Tezza, Caprio, Pinto, et al. (2020) | V | M | virtual drone movement + engagement score | finish race as fast as possible (LT) | 3N | | | | | U | | | | | ✓ | |
| *[185]* | van de Laar et al. (2012) | V | M | brain signals + score increase | score as many points as possible (LT) | | | B | P | ✓ | U | | | | | | |
| *[186]* | Vourvopoulos & Badia (2016) | V, A | M, HMD | virtual hand moves lever + door opens + accompanying sounds | rotate lever to open garage door (ST) | 1H | | | | | | ✓ | | | | | |
| [45] | Vourvopoulos et al. (2016) | V, A, H | HMD | virtual hand rows + user hand vibrates | row a boat (ST) to indicated flags to collect as many flags as possible (LT) | 1H | ✓ | | P | ✓ | D | ✓ | | | | | |
| [30] | Vourvopoulos et al. (2017) | V | M | ball moves | move through maze (ST) to reach maze exit (LT) | | ✓ | I | U | | | | ✓ | | | | |
| *[132]* | Vourvopoulos et al. (2019) | V, A, H | HMD | virtual hand rows + user hand vibrates | row a boat (ST) to indicated flags to collect as many flags as possible (LT) | 1H | ✓ | | P | ✓ | D | ✓ | | | | | |
| *[187]* | Wang et al. (2012) | V | M | avatar walks | move to next NPC (ST) to complete all stops (LT) | 3H | ✓ | N | P | | | | | | | | |
| *[125]* | Wang et al. (2019) | V | M | Tetris block moves | move block to prevent it from stacking and place it to form lines (ST) to maximize survival time and score (LT) | | ✓ | N | P | ✓ | | | ✓ | | | | |
| *[188]* | Xie et al. (2022) | V, A | M, HMD | varied, virtual arms and characters moving + audio | varied | 1H | ✓ | B | | | ✓ | | ✓ | | | | |
| *[126]* | Xu et al. (2020) | V | n/a | virtual robot arm moves | n/a | 3N | | | | | | | | | | | |
| *[189]* | Yang et al. (2018) | V | M | platform rotation + planet rolls + waveforms plot + cartoon face | move platform to balance planet sphere (ST) to maximize score (LT) | | ✓ | | | ✓ | | | ✓ | | | ✓ | |
| *[190]* | Ziadeh et al. (2021) | V, A | HMD | virtual hand pops balloon or buzzer sound | grab balloon to pop it (ST) | 1H | | | | | | | | | | | |

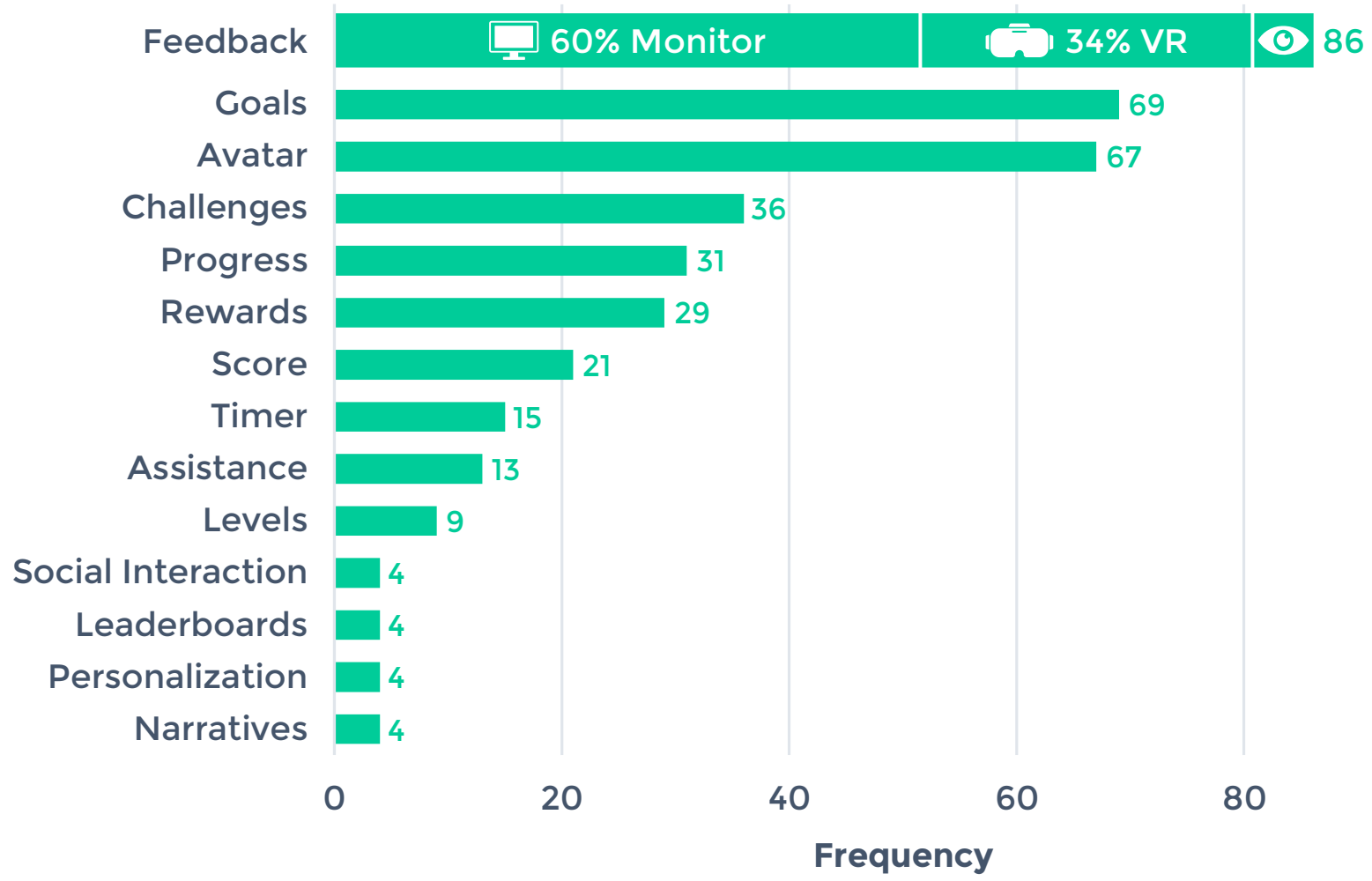

*Figure 5.* Number of studies implementing each element and the proportion of each platform used to deliver visual feedback (i.e., display monitor, virtual reality, or the real world).

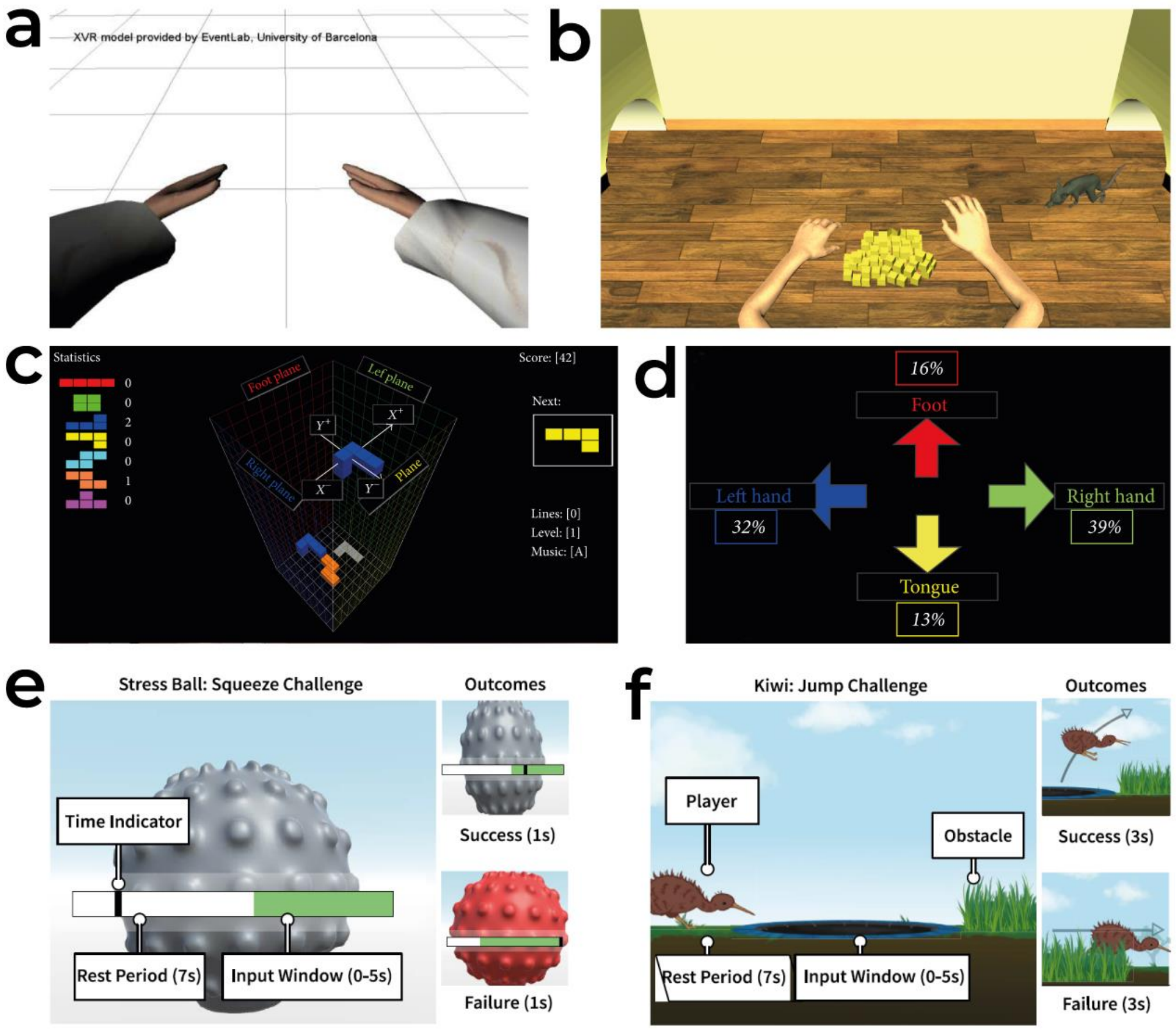

*Figure 6.* Examples of gamified MI-BCI protocols: (a) simple VR feedback [109], (b) narrative-based goal-oriented game [110], (c) 3D Tetris [111], (d) 2D Screen Game [111], (e) simple ball squeeze task [112], and (f) narrative-based kiwi runner [112]. Reproduced with permission from Springer Nature, Frontiers, Hindawi, and original authors.

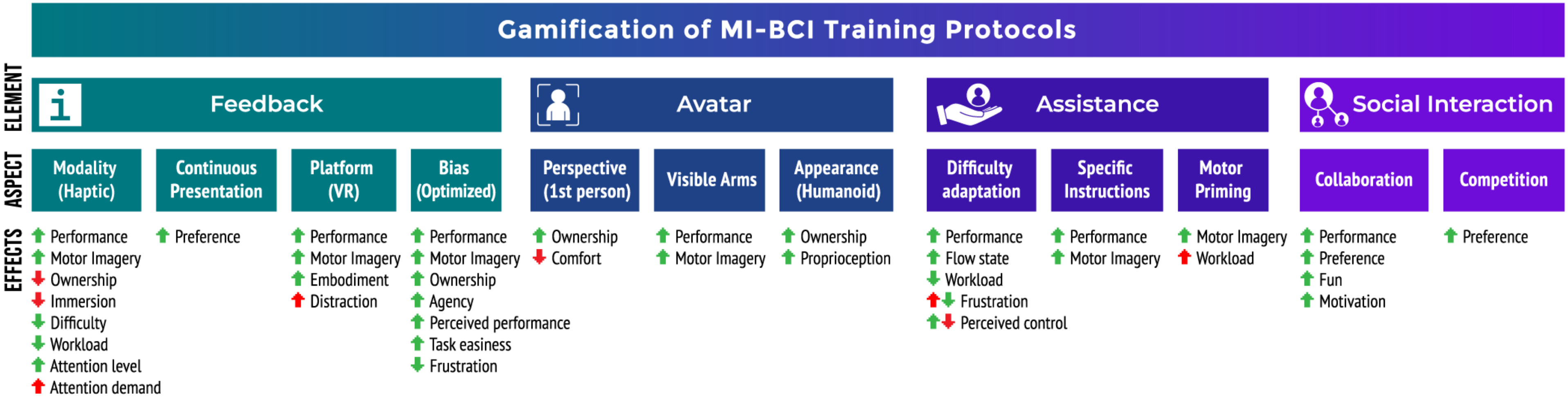

*Figure 7.* Summary of the impact of gamification on MI-BCI training. The arrows indicate reported positive (green) and negative (red) effects of individual game elements on user performance, experience, and motor imagery. Null results, including those of the elements score and timer, are excluded. The remaining eight elements were not studied empirically in any of the 86 papers included in this review.

## 4.2. Feedback

Feedback is inherent to both gamification and MI-BCI training. This commonality offers a unique opportunity to gamify an existing aspect of BCI training. As real-time BCI use was a criterion of this study, feedback was present in all 86 included studies. However, only 24 of these studies reported on the effects of various feedback aspects on user outcomes. While the feedback given to users across studies was unique in terms of content, patterns emerged in the sensory modalities and platforms through which feedback was given (Table 3).

### *4.2.1. Feedback Modality*

Feedback was overwhelmingly given visually, commonly as objects or avatars moving in a virtual environment. Some studies also provided auditory feedback, such as a buzzer sound when performing the wrong movement or sound effects accompanying the visual feedback, e.g., a balloon popping [190]. In contrast, haptic feedback, which can either be tactile or kinesthetic, was used least often. Tactile feedback was delivered either mechanically, for instance through a vibrating wearable [114], or electrically by stimulating muscles [135]. Kinesthetic feedback was given to users by having external devices force their body movements, such as a robot orthosis making the user's hand close/open [109].

Of these modalities, particularly haptic feedback was studied in terms of its effectiveness on various user outcomes. In Petrushin et al. [176], users found it easier to control a robot navigating a room when they received kinesthetic-tactile feedback where their finger pushed a button, mimicking a computer mouse click. While the addition of this feedback demanded more attention, users achieved significantly higher BCI accuracies than without the feedback. This suggested that tactile feedback could improve user performance and help differentiate MI responses. Similarly, adding tactile feedback to visual feedback in VR in the form of functional electrical stimulation (FES) resulted in better BCI performance and more intense motor cortex activity during MI [178]. Achanccaray et al. [135] also confirmed this by reporting better BCI performance for grasping MI when providing FES on top of VR-based visual feedback. Additionally, their results suggested that FES could facilitate the MI task by recruiting the premotor cortex during MI preparation. While not significantly different, the multimodal feedback (VR+FES) was generally rated higher on attention paid, and lower on body ownership, immersion, and task difficulty.

Rather than examining the effect of one (additional) modality as above, some studies compared individual modalities in separate conditions. While VR feedback and tactile feedback through FES resulted in similar BCI accuracies, the latter incited larger EEG amplitudes during MI [183]. In two studies, kinesthetic feedback of a robotic device moving the user's hand resulted in significantly better BCI performance than visual feedback of a virtual arm or bar [146,179]. In addition, the mental workload was lower with kinesthetic feedback and it enhanced MI cortical activity through more pronounced ERD/ERS [146,179]. In contrast, solely auditory feedback did not differ from solely visual feedback in terms of BCI performance [166]. Nevertheless, all users reached effective control with auditory feedback and with it users improved more over the sessions than with visual feedback, suggesting auditory feedback to be a suitable alternative to visual feedback. Similarly, Kreilinger et al. [120] added sound effects (e.g., beeps) to existing visual feedback to indicate successful and unsuccessful MI actions, but found no impact on online task performance. Finally, instead of examining single modalities, Arpaia, Coyle, et al. [138] compared visual + haptic feedback to no feedback at all, revealing that multimodal feedback improved BCI performance.

#### *4.2.2. Delivery Platform*

Looking into how visual feedback was delivered to users (Fig. 5), 59 studies utilized a normal display monitor or screen. VR on the other hand, was used in 34 studies, mainly through a head-mounted display (HMD). The least used was real-world feedback, in six studies, which involved feedback that was provided directly to the user. For example, users could use the MI-BCI to control a real wheelchair to navigate a room [163] or fly a drone in front of them to a target height [131].

With normal monitors being the most frequent, it is of interest to discern whether they are also the most effective platform for feedback delivery. This is, however, difficult to determine because some studies compared these platforms with distinct visual content. For instance, compared to a 2D abstract bar on a monitor akin to the classic Graz protocol (Fig. 2), VR feedback of a virtual arm resulted in lower error rates [152,161,174]. However, users also found the VR feedback to be more distracting when the wrong arm moved, that is, negative feedback. Škola and Liarokapis [182] corroborated this positive effect of VR on BCI performance using similar conditions, but they also found that the sense of agency and ownership were rated similarly in both conditions. Likewise, Xie et al. [188] found that VR scenes of virtual arms resulted in higher BCI performance compared to dynamic (animation and games) and static scenes (pictures and text) on a monitor. In addition, MI neural activity was stronger for VR scenes as indicated by more obvious ERD/ERS. Because each of these studies compared a 3D virtual environment presented through VR to a 2D abstract bar presented through a monitor, it is hard to disassociate the effects of the VR platform from those of the feedback content.

Several other studies can shed light on this. Mahmood et al. [165] presented the same visual feedback through an HMD and monitor (Table 3), revealing that the HMD improved BCI performance by helping users visualize MI better. Similar results were found by Achanccaray et al. [134] who presented a virtual environment through both an HMD and a monitor. In contrast, Juliano et al. [156] found no significant difference in performance when comparing the same visual feedback of a virtual arm presented through an HMD or on a screen. There were also no differences between the two conditions in terms of simulator sickness, presence, and sensorimotor activity, but the HMD did incite higher embodiment. Furthermore, only with an HMD was there a positive correlation between embodiment and performance, altogether suggesting that HMD feedback could increase BCI performance by increasing embodiment.

Extending this scope, Blanco-Mora et al. [143] compared three conditions: ‘immersive VR’ (avatar rowing in a virtual environment through an HMD), ‘non-immersive VR’ (same feedback on a monitor), and the classic Graz protocol (bar feedback on a monitor). While the VR conditions outperformed the classic Graz protocol, there was no performance difference between immersive and non-immersive VR. This suggests that the visual content of the feedback might affect training rather than the delivery method. However, this is not consistent with Vourvopoulos and Badia [186], where both a virtual environment and the classic Graz bar were presented through an HMD. No significant differences were found in BCI performance or motor imagery ability, but VR did elicit stronger motor network activity that could help with motor rehabilitation.

One possible explanation for these contrasting results is that the VR platform (*how* feedback is delivered) interacted with the visual content (*what* is delivered). This interaction can be investigated in future work with a controlled two-by-two factorial design to discern the true impact of the delivery platform relative to the feedback content on user experience and performance.

### *4.2.3. Feedback Content*

#### ***4.2.3.1. Feedback Congruency***

The extensive visual feedback presented to users was often in the form of virtual arms and avatars (detailed further in section 4.4) naturally coupled to the user's imagery of movements, e.g., imagining a right-hand grasp to activate a virtual right-hand grasp [146]. Such anatomical or spatial congruency that mimics the user's mental imagery of movement was mentioned to be of importance to the user's embodiment [112,158,190].

Also of importance is that the feedback is timed with minimal delay, that is, temporally congruent with the user's motor imagery. Kjeldsen et al. [158] studied such temporal congruency of visual feedback. They compared feedback where a virtual hand moved continuously without delay based on the classifier's output strength, to discrete feedback where the whole movement was performed at once when an output threshold was reached. Users preferred continuous feedback, while discrete feedback was rated as not useful, even making users give up because they did not see a hand moving in failed MI attempts. Of note, only for discrete feedback did agency over the virtual hand positively correlate with BCI performance. This is relevant to users with poor BCI control, as discrete feedback would lower feelings of control over BCI-controlled movements. Therefore, continuous feedback might be more beneficial to MI-BCI training.

#### ***4.2.3.2. Feedback Bias***

Some of the aforementioned studies found that negative feedback can be distracting because of the mismatch between the user's intended movement and the visual feedback, e.g., when a left virtual hand grasps while the user imagined a right-hand grasp. This issue could be alleviated by presenting only positive feedback, that is, when the user's intended movement matches the cued movement [152,161,174]. This is corroborated by Alimardani et al. [44], where users were given either (1) no feedback at all, (2) only positive feedback, or (3) both positive and negative feedback when they controlled humanlike robot hands with a MI-BCI. Users rated agency and ownership over the robot hands higher when given only positive feedback. A second experiment took this further by giving positive feedback regardless of whether the user's classified output was correct. Such positively biased feedback resulted in higher MI learning and more classifiable MI patterns, relative to non-biased feedback, solely positive feedback, and negatively biased feedback. Similarly, Alchalabi et al. [137] found that positively biased feedback improved

BCI performance, event-related synchronization, body ownership, agency, perceived performance, and easiness of tasks, compared to regular feedback.

Considering biased feedback impacts user outcomes, how much bias is then optimal? Hougaard et al. [112] gave different percentages of guaranteed positive feedback. The positive feedback rate predicted user agency and frustration, with higher rates increasing agency while decreasing frustration. Contrary to Alimardani et al. [44], Mladenović et al. [168] found negatively biased feedback to be most beneficial for user performance and motor learning, compared to positively biased and non-biased feedback. However, these benefits only lasted for one session and became impeding for learning by the second session, even more so for low-performing users. Of note, biased feedback interacted with several user factors such as workload, suggesting it should be tailored to the user. Finally, the authors noted that the inconsistent findings across studies on feedback bias could be explained by the variety in interface designs and approaches to feedback.

### 4.3. Goals

Table 3 shows the goal(s) defined in each study's protocol: the desired outcome of the user's actions within the environment. Actions could be performed with the BCI to complete a goal either per trial (short-term goals) or in the longer term for the whole session (long-term goals), the latter requiring a series of actions. Overall, 69 studies (80%) had some form of goal, with 59 having at least a short-term goal per trial and 44 studies a long-term goal. Looking further into the goal descriptions, many studies had trials where the goal was to interact with an object within a virtual environment. For instance, users had to control virtual arms to catch falling fruit [130], grab balloons to pop them [158,190], or press buttons [172,182].

Similarly, long-term goals also showed a pattern, where the goal of the whole experimental run or session was often to maximize a score. This was typically either in the form of earning as many points as possible by performing correct MI actions [140,157,162,169,185,189], or conversely minimizing time spent on the task, such as the lap time in racing games [128]. A final noticeable trend were target-oriented goals, where the users had to move their avatar to an indicated position, such as a maze exit or the finish line of a track [30,116,148].

Despite so many studies implementing goals in different forms, neither the presence of goals nor their specific implementation has been studied empirically. Therefore, its effects on user performance and experience in MI-BCI training remain unclear, requiring future research.

**4.4. Avatar**

While technically a type of visual feedback, avatars serve a more specific function: they visually represent users and enable them to interact with the environment. Out of the 86 studies, 67 (78%) had the user control an avatar using their motor imagery. Most commonly, in 37 studies the avatar was controlled from the first-person perspective, where users saw the environment from the viewpoint of the character's eyes. From this perspective, the avatar was almost always human, often referred to as an 'embodied' avatar when combined with VR (Fig. 8a). Only 2 studies used a non-human appearance from the first-person perspective (a virtual and physical robot). In three other studies, the avatar had no appearance at all, but it was evident from the description that the user moved through the virtual environment as an avatar, e.g., embodying Little Red Riding Hood [160]. In contrast, the third-person perspective, where users saw the entire avatar in front of them, was used in 32 studies: 25 times with a non-human avatar, such as animals and robots (Fig. 8b), while only seven times with a human character (Fig. 8c).

The total number of avatars was 69 across these 67 studies because two studies employed an avatar from different perspectives to examine the effectiveness of this aspect. Comparing first- and third-person perspectives using a virtual robot avatar, Davis et al. [150] found no significant difference in BCI performance. However, the first-person perspective was slightly less comfortable due to continuously looking down, which might have influenced the subject's MI and subsequent performance. Another study comparing the two perspectives using virtual human arms found similar results [139]: while a first-person perspective induced more body ownership, there was no difference in user agency and BCI performance. This suggested that neither the perspective nor body ownership induced by said perspective influences MI-BCI control.

Regarding the avatar itself, Choi et al. [148] found that the presence of virtual hands in a first-person perspective resulted in better user performance than without visible hands. Furthermore, the visible hands induced more discriminable brain signals and MI-related cortical activity. As BCI accuracy was positively correlated with embodiment and presence scores only when virtual hands were visible, a possible mechanism by which visible avatars contribute to BCI control may be through embodiable feedback. Concerning avatar appearance, Ziadeh et al. [190] revealed that users felt higher ownership and proprioception when grasping an object with virtual human hands than with abstract blocks. Agency and frustration were however rated similar, and MI performance did not differ between the two appearances.

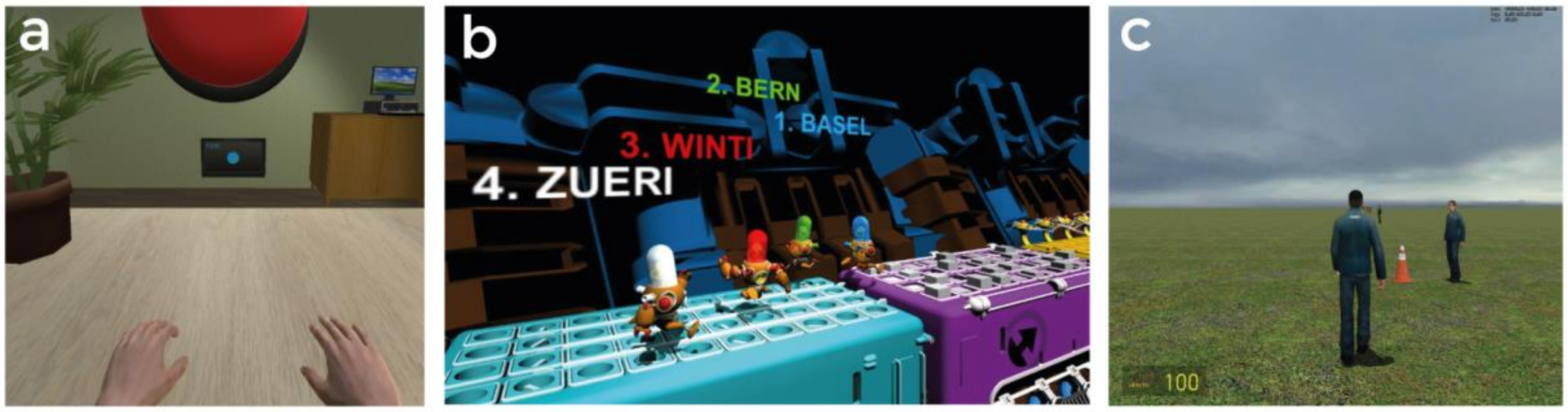

*Figure 8.* Examples of avatars controlled with MI-BCI: (a) humanlike arms from the first-person perspective [182], (b) robots from the third-person perspective [173], and (c) a human character from the third-person perspective [187]. Reproduced with permission from Elsevier, IOP Publishing, and Frontiers.

**4.5. Challenges**

Thirty-six studies (42%) implemented a challenge for the player to overcome. While these varied significantly as challenges were tailored to the study design and goal, some patterns could be identified. For instance, several studies implemented some kind of obstacle for the player to avoid [112,118,120,122,139,140,166,173,175], e.g., controlling a car to avoid incoming barriers while collecting coins [120].

Other challenges relied on a user's ability to control the BCI system, requiring activation of MI commands with appropriate timing, strength, or duration [45,119,132,142–144,155,163,164,187–189]. This for example involved making a penguin avatar jump on time to catch a fish [162], or stopping one's avatar as close as possible to a target and thus avoiding under- or overshooting [142,144,189]. These challenges require the BCI to detect not only the specific movement the user imagined, but also the strength of their imagery. Furthermore, relying on user timing requires constant classification of MI, that is, a self-paced BCI. In comparison to these complex implementations, a simpler way to challenge users was to make them compete against each other [145,173,175,177], although in return the study would require multiple participants using a BCI simultaneously (see 4.10 as well).

The challenges mentioned so far relied mainly on the user's ability to complete requested actions. In contrast to these 'behavioral' challenges, some studies implemented challenges of a more cognitive nature, requiring the player to think about the best course of action. This included solving puzzles [111,121,125], or deciding on the best path to escape a maze [30,116]. A final way challenges were implemented was by setting explicit time limits that pressure the user

[45,111,119,125,131,132,143,160,163,164,168–170,187], which was common when the overall goal was to gather as many points as possible.

None of the 86 studies empirically examined the effects of challenges on MI-BCI training outcomes. However, McCreadie et al. [166] found a positive correlation between online BCI performance and the perceived challenge, measured subjectively from users. It might be possible then for more challenging training protocols to improve MI-BCI performance, but this warrants future investigation into the causal effect of challenges on MI-BCI performance.

### 4.6. Rewards

Any incentive provided to the user for performing actions successfully can be constituted as a reward, which a total of 29 studies (34%) did. The most common type of reward was points given for performing a correct (MI) command within a trial and consequently achieving short-term goals such as collecting coins. These points increased the user's score and could visually be presented in thematically fitting graphics, e.g., stars [117]. Less commonly, points were given for achieving longer-term goals such as navigating to flags [45,132,143] or forming lines in Tetris [111,125], each requiring a series of correct MI actions. Interestingly, all studies that implemented rewards also implemented goals, highlighting that rewards are given for specifically targeted actions that achieve in-game objectives.

Some studies gave users rewards unique to their protocol design: puzzles pieces to complete a puzzle [133], fish to reel in when playing as a fisherman [155], a speed boost in a racing game [122,173,175], and unlocking the next level in a maze game [30]. One protocol adjusted the rewards in proportion to the difficulty of the challenge, where collecting objects on the far edge of the screen through stronger MI resulted in more points than objects closer to the center [168]. Similarly, users could receive more points by performing multiple correct actions in a row, resulting in combo streaks and multipliers [125,142,144].

In addition to rewarding correct actions, some studies also penalized incorrect actions resulting from wrong motor imagery (classification). For instance, the user's score was reduced by a point [120], or users had a limited number of actions, compelling them to avoid false-positive MI activity that used up their actions if timed wrongly [117]. The penguin race by Leeb et al. [162] also penalized users for false positives: wrong MI (classification) would result in accidental, wrongly timed jumps, which would make jumping on time for the next trial (i.e., next fish to collect) difficult. Similarly, in the Cybathlon racing game, misplaced commands slowed down, rather than sped up, the user's avatar [122,173,175]. Finally, in Vourvopoulos et al. [30], mazes had open gaps where the user's ball could fall off, resetting their position.

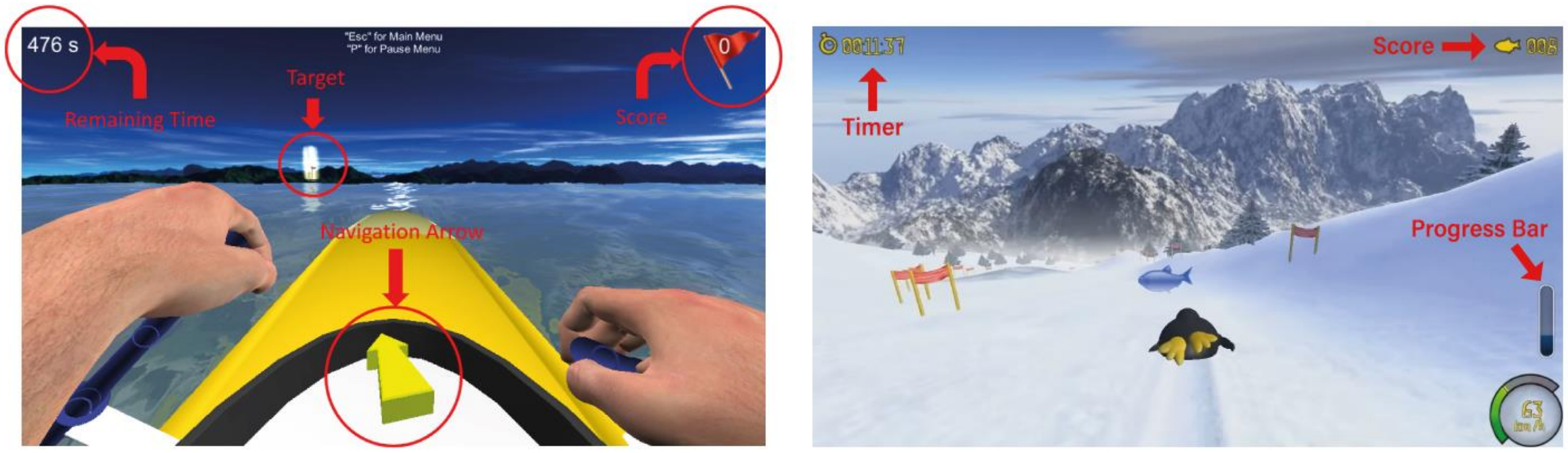

*Figure 9.* Examples of visual indicators used in gamified MI-BCI protocols from (left) NeuRow [183] and (right) TuxRacer [156]. Reproduced with permission from Springer Nature.

Despite this diverse implementation of rewards, it is unclear how they influence MI-BCI training outcomes as none of the studies examined and reported on the individual effects of this game element.

### 4.7. Visual Indicators: Progress, Score and Timer

In-game information can be communicated visually through on-screen indicators, such as numbers and abstract shapes. Figure 9 shows examples of three common types of visual indicators implemented in MI-BCI protocols. Firstly, 31 studies (36%) indicated progress to their users. Of these, the majority used a simple number representing the trial or step the user was at in the session. Other studies used a progress bar, most often to indicate the avatar's position relative to reaching a long-term goal such as the end of a racetrack. However, a few times the bar indicated a user's progress within a trial, reflecting their MI activity or how close they were to scoring a point, akin to the traditional bar feedback in a Graz protocol trial (Fig. 2). Finally, in a few studies the progress was not explicitly indicated through a number or a bar, but only implicitly visible within the environment. These were, for instance, a puzzle being filled in with pieces earned through correct MI actions [133], or a word being filled in with letters in a BCI version of the classic game Hangman [154].

The second type of visual indicator was a score, which 21 (24%) studies implemented to visually and quantitatively inform the user of their performance. In almost all studies, this was achieved through a simple number on the user interface that was constantly shown while users performed the MI tasks. However, two studies displayed the score not constantly, but after each trial or run [33,124]. Unique to Škola et al. [33], the score was not shown on a user interface around the screen, but on a small display on an object within the virtual environment.

The final visual indicator type was a timer, which visually counts the in-game time down or up. In MI-BCI training, fifteen studies (17%) implemented some form of timer. Almost all timers were in digital format, counting the minutes and/or seconds. Most often the timer counted up starting from zero, to indicate the elapsed trial or session time. However, in four studies the timer counted down, explicitly communicating the time limit of the task to the user.

To understand the effectiveness of these visual indicators on user outcomes, Castro-Cros et al. [110] compared a gamified protocol without any score and timer to one where the two visual indicators were shown either constantly, or intermittently (after every 10 trials). No differences were found in BCI performance between the three conditions, suggesting that neither the element nor the frequency of its presentation matters. Because this study still implemented the two visual indicators together, these findings should be taken with caution. For progress, it is unclear how indicating it to the users impacts their training outcomes as the element was not studied empirically in any study.

### 4.8. Assistance

Thirteen studies (15%) assisted the user in some manner. The drone racing game by Tezza et al. [128] provided users with a tutorial to teach the rules and controls. Another BCI drone game changed the activation threshold of MI commands based on user performance [131]. Such a difficulty adjustment was proposed to engage and motivate users, and was also implemented by Badia et al. [115]. Similarly, studies have given artificial boosts to overcome or automatically destroy an obstacle when users could not achieve it with their MI [33,139]. Finally, in NeuRow, a VR-based rowing game, users were stopped from deviating too much from the target they had to navigate to [45,132,143].

Several other studies both implemented and studied the effectiveness of assistance. These can be distinguished into assistance provided before the user's interaction with the BCI system and those provided during their interaction. For the former, Battison et al. [141] provided users with three types of instructions: non-specific (to imagine a hand movement), specific (to imagine squeezing stress balls), and participant-specific (to imagine hand movements they frequently perform). More specific instructions, particularly participant-specific ones, led to less training time, greater mu power, and better BCI performance. Also taking place before user-BCI interaction, Vourvopoulos and Badia [186] had users physically execute movements before imagining them in a VR task. Such motor priming resulted in stronger motor network activation during motor imagery, while users also rated the condition with motor priming to be more demanding in terms of workload. Nevertheless, there were no significant differences in BCI performance or reported kinesthetic imagery ability compared to no priming.

In contrast to the above studies, several other studies provided assistance during the user's interaction. Mladenović et al. [169] adapted the protocol's difficulty by adjusting the classifier output to the user performance, where the avatar moved more in the desired direction when the user performed poorly, and less extreme or even into the opposite direction when the user performed well. This resulted in higher reported cognitive control, a dimension of user flow state, compared to a non-adaptive condition. However, there was no difference in online BCI performance. Similarly, Saeedi et al. [124] studied online adaptive assistance, where users either had a fixed time limit to deliver their command or the time limit adapted if they took too long. Adaptive assistance resulted in significantly higher success rates and reduced the task workload.

Another way to adjust the difficulty is through performance accommodation mechanisms: in-game mechanics that alter the user's output to accommodate for their performance, thereby increasing their enjoyment and motivation to continue the game [155]. In a game where the user had to reel in fish using MI-BCI, three such mechanisms could help them in the trials: (1) 'augmented success' made each correct MI command count as two successful actions (similar to a power-up or boost), (2) 'mitigated failure' turned unsuccessful MI attempts into neutral feedback by neither catching nor letting the fish escape (reducing the penalty for failed commands), and (3) 'input override' turned unsuccessful MI attempts into positive feedback by having a non-player character catch the fish for the user (replacing failed commands). Whereas the latter two were found to be frustrating and reduce perceived control, augmented success did not. Furthermore, users with low BCI control had higher perceptions of control with the mechanisms, while the perceived control of users with high BCI control was lower. This suggests that the performance of proficient users was obstructed by assistance.

### 4.9. Levels

Another way for users to experience a change in difficulty is through levels. While difficulty adaptation sets a continuously varying difficulty for each user depending on their performance at any time, levels have a fixed difficulty that each user goes through in a stepwise manner as they progress through the game. Nine studies (10%) employed this element in their training protocol, most commonly in the form of separate environments or maps the user progressed through. These different levels typically increased in difficulty by changing an aspect of the environment [30,162,166,170]. For instance, users could advance through a series of mazes where each maze had more intersections or dead ends than the previous level [30,170].

Levels did not always change in terms of both environment and difficulty. In some studies, levels could be unlocked as new environments by progressing or using points, without any mention of difficulty changes [157,188]. Conversely, levels did not always consist of different environments; users instead played within the same, continuous environment, with their level indicated through a simple number or text on the interface [111,125,189]. In these studies, levels indicated the game difficulty, which users progressed through if they performed well or reached set goals. A familiar example of this was Tetris: the level goes up if users clear a certain number of lines, which in turn increases the falling speed of blocks. In an MI-BCI adaptation of this game [125], this meant users had less time to perform the MI task, thus increasing the difficulty of the game.

Some of these studies implemented levels to keep the performance dynamic and engagement high [30] or to give users new environments to explore [157]. However, the actual effectiveness of levels on BCI performance and user experience is unclear as there were no reported effects among the included studies.

### 4.10. Social interaction

Implementing social interaction in MI-BCI training requires a multi-user BCI setup. Four studies (5%) had such a setup where multiple users simultaneously used the MI-BCI to compete or collaborate on the task. In the Cybathlon Race [173,175], two to four users competed in a virtual race where they had to perform correct actions to avoid obstacles and control their robot character's speed. Similarly, Pradhapan et al. [177] had two users compete in a real-world race with toy cars remote-controlled through MI-BCI. The final study by Bonnet et al. [145] included both competition and collaboration in a simplified soccer game. Two users moved a ball in the center of the screen either to the designated goal line through their combined motor imagery (collaboration) or to the opponent's goal line through their individual motor imagery (competition).

Bonnet et al. [145] was also the only study to empirically study the effects of competition and collaboration on BCI performance and user experience. They first compared a single-user condition to a collaborative two-user condition and found no significant difference in BCI performance. However, dividing users further showed that high-performing users performed better in the collaborative condition, while low performers showed no differences between conditions. Furthermore, collaborating was rated as more fun and motivating, although at the same time more users preferred playing solo if they had the choice. Adding a competitive condition in a follow-up experiment, there were again no differences in BCI performance between playing solo, collaborating, or competing. Once more, the high performers performed best when collaborating while low performers performed best when solo, although differences were not significant this time. Similar results were found for the user experience in this second experiment: the two multi-user conditions were rated as more fun and motivating, and preferred over solo play this time, but these differences were not statistically significant. It should be noted that this follow-up experiment with competition had only eight subjects, which might have been too small of a sample size to detect effects between the three conditions. Altogether, these findings indicate that social interaction is at least beneficial for BCI users with good control. Nevertheless, with results from only one study, more evidence is warranted to support the effectiveness of this game element.

**4.11. Leaderboard**

Related to social interactions are leaderboards, which show user rankings relative to each other and were used in four studies (5%). In the BrainRunners game of the 2016 Cybathlon [173,175], leaderboards were implemented in the form of participants' live positions on the track. Thus, not only could participants see their own progress as mentioned before, but they also saw the progress of their opponents, allowing them to rank themselves relative to others while playing. In two other studies [128,157], leaderboards were implemented in a more classic form of a visual ranking that users could view to check the performance of their and other users' previous runs. While the aim of leaderboards in Kapitonova et al. [157] was to motivate players and increase game replayability, the actual effectiveness of leaderboards was not studied in comparison to a control condition, nor was it by any other study.

### 4.12. Personalization

Four studies (5%) implemented personalization, where users were allowed to customize an aspect of their experience. A drone racing game [128,129] allowed users to select their drone avatar, the racing track, and auditory (i.e., sound effects) and visual distractions (i.e., fireworks, flags). In another version of the same game [184], users could also choose to display their engagement score, measured using a passive BCI. Finally, personalization was also implemented by letting users change the environment through a 2D background image of choice, the music played, and even the game's difficulty [189]. The last is akin to how video games let users choose their preferred difficulty level at the start of a game, as opposed to the system adapting the difficulty (see Assistance) or the user having to progress through a set difficulty level (see Levels). Among these few studies implementing personalization, there were no reported effects on any MI-BCI training outcomes, hence this element remains understudied.

### 4.13. Narratives

While goals can guide user actions, narratives can give context and meaning to actions to further engage the user. This element was implemented by four studies (5%) in the field of MI-BCI training. Namely, protecting a pile of cheese from a rat using virtual arms (Castro-Cros et al., 2020), protecting Earth from incoming asteroids by shooting them down [33], Little Red Riding Hood journeying home in the dark while avoiding dangers [160], and a mother kiwi bird running home to protect their eggs from a predator [112]. Overall, the narratives implemented were simple stories with few characters and events, where the overarching theme was to either protect or navigate to a target. These few studies implemented narratives in combination with other elements, without a comparison or control condition, therefore it is not possible to ascertain the individual effectiveness of narratives on user experience and performance in MI-BCI training.

## 5. Discussion

The aim of this systematic review was to identify and evaluate the use of gamification in motor imagery brain-computer interface training. Through a systematic literature search, we identified 86 studies from the past decade that employed gamified protocols for MI-BCI training. Overall, gamification seems to have positive impacts on both the experience and performance of users controlling a brain-computer interface with their motor imagery (Fig. 7). This is in line with Gao et al. [40], who found gamification to be viable and positively reported on for MI-BCI training. Unlike their work though, this review identified the implementation of individual game elements in MI-BCI training protocols. We found many studies to use control conditions, revealing empirical support for the effectiveness of gamification and specific game elements on both user performance and experience. The aggregated results show that gamification is most often implemented through gamified feedback, goals, and avatars. This differs from other fields such as education, where points, badges, and narratives were most common [35,37,67,81]. Henceforth, we discuss the implementation and effectiveness of each element in more detail to establish a framework with concrete recommendations for gamifying future MI-BCI training protocols.

### 5.1. Effective Recommendations for Gamification

Feedback was naturally the most frequent element as it is structural to the real-time implementation of the MI task and user-BCI interaction. This aspect was often gamified by showing interactive objects and avatars in a virtual environment. Such feedback might be more comprehensible to the average user than an abstract bar or recorded brain signals, in line with Scherer et al.'s [107] suggestions. Furthermore, any changes in the virtual environment are best shown continuously, in synchrony with the strength of the user's motor imagery [158]. In comparison to visual feedback, haptic feedback was more beneficial for both user performance and experience, even inciting cortical activity during motor imagery. However, it is more costly and difficult to implement, thus researchers should consider the trade-offs. Due care must also be exercised when implementing auditory feedback such as sound effects. While auditory feedback alone can be a suitable alternative to visual feedback [166], adding it to visual feedback has no clear benefits [120]. On the contrary, sounds may even be detrimental to the training process as background music worsened BCI performance in one of the included studies [169].

Visual feedback was most frequently delivered through standard monitors. However, they were not the most beneficial to training, as VR resulted in better user performance, experience, and MI-related brain activity. This substantiates the previous literature that highlights the potential of VR as a training platform for BCI [96,101,103–105]. Not all evidence converges though, as it is still unclear from the results how much the manner of delivery (e.g., VR or monitor) plays a role relative to the visual content that is delivered (e.g., 3D virtual environment), and how these factors interact [134,143,165]. In terms of feedback bias, results generally suggest leaving out negative feedback, thus only showing feedback for successful MI trials. Some results even point to showing positive sham feedback for unsuccessful outputs, subsequently 'biasing' the feedback, but support for this is mixed, hence requiring further investigation. The mixed findings might be partly explained by user factors, as feedback bias was found to interact with users' personality, workload, anxiety, and self-control [168].

Many studies had users control an avatar with virtual arms, perhaps because such visually congruent mapping of imagined movements and performed actions in the environment can be beneficial to user performance and experience [22,23,191,192]. The results suggest that the presence of virtual arms increases BCI performance and MI-related brain activity [148]. However, avatar appearance does not influence BCI performance, despite human hands inciting more ownership [190]. Likewise, a first-person perspective incited more ownership than a third-person perspective, but it did not influence user performance or agency [139,150]. Taken together, users might benefit most from controlling a human avatar with limbs from a first-person perspective. As embodiment only correlates with performance when virtual arms are visible [148], such an avatar might induce stronger effects when presented in VR, which has been shown to increase embodiment [100,105].

Another element whose effectiveness was elucidated is assistance, which encompasses a variety of techniques that can help users before and during their interaction with the BCI system. Of note during the interaction is the adaptation of difficulty to the user's skill, for instance by giving them more time to output the correct MI command or amplifying the output if the user performs poorly. This is important as too easy tasks bore users, while too difficult tasks make them quit [91]. Results show that difficulty adaptation can improve BCI performance, reduce user workload, and augment a flow state, which is in line with cognitive research and game design [193,194]. These findings lend empirical support to the suggestions made by previous BCI reviews [39,107]. Nevertheless, caution should be taken as too much assistance, such as completely overriding the user's commands [155], can be frustrating and reduce agency over the BCI system.

Users can be further assisted with an introductory period that should be treated with the same care as the main task interaction [22]. This introductory period can be implemented in two ways before the interaction. Firstly, giving specific instructions about imagined movements can help modulate sensorimotor activity and improve BCI performance [141]. Secondly, physically executing movements before MI, known as motor priming, can accentuate motor-related brain activity during the MI task [186]. We also propose to gamify instructions by embedding them into a tutorial, which only one study included [128], despite being essential in game design for teaching game rules and controls [72,86]. This could remove possible inconsistency in verbal instructions and provide users with more implicit guidance by making them consider instructions as part of the game [195].

Similar implicit guidance was already implemented by some studies as gamified cues (Table S3 in Supplementary Materials). Specifically, the environment cued the limb movement users had to imagine by putting an interactive object at the corresponding position, e.g., coins to collect on the left side of the avatar. This subtly implies the best action to take using the appropriate MI command rather than directly instructing the action. In video games, the environment and objects within it commonly inform players of possible actions for gameplay [196–198]. Such non-controlling instructions can enhance user autonomy and motivation by creating an illusion of choice [199–201]. Previous BCI reviews similarly recommended embedding stimuli in the game environment and guiding users with objects [98,102], but this line of reasoning still needs to be confirmed in a future study.

Previous work on BCI games also recommended multi-user training protocols where two or more users collaborate within the same game [102]. Our review reveals that such social interaction can help MI-BCI training. Collaboration and competition are more fun and motivating than solo play and are preferred by a subset of users. Additionally, it can improve BCI performance for users that already perform well. However, these findings are only supported by one study as social interaction is rarely implemented in MI-BCI training. This could be due to the difficulties that come with setting up a multi-user BCI system. A less costly alternative to human social interaction could be to use virtual agents or non-player characters (NPCs). Extensive research is still underway to make virtual agents as realistic as possible, both appearance- and behavior-wise [202–205], but they have already been capable of creating a realistic sense of human interaction [206–208].

Gamification can also allow MI-BCI protocols to communicate the user's performance through visual indicators, rather than burdening them with guesses of their performance [107]. Namely, a score can be displayed to users to self-monitor their performance on the task, e.g., the number of successful MI actions. This number is mostly displayed on the user interface layered on top of the virtual environment but can also be embedded as part of the environment as in Škola et al. [33]; a more immersive concept from game design termed diegetic user interfaces [209,210]. Due to its simplicity, scores may be more comprehensible to the average user than BCI accuracy. Likewise, users can be made aware of how much time has passed or is left through a timer, which can pressure the user to perform actions and complete goals [85,89]. Although these visual indicators did not impact user performance [110], their effect on user experience is yet to be examined and thus requires further investigation. Compared to scores, timers might be easier to implement as time limits are already inherent to the experimental nature of MI-BCI training, per trial and per session.

Other information that can be communicated concerns the user's progression in the environment, which allows them to locate themselves relative to the goals [88,89]. This review did not encounter any evidence for the effectiveness of progress, whether communicated through a simple number or an abstract bar. Similarly, levels (of increased difficulty) that users could advance through were also neglected. Vourvopoulos et al. [30] implemented such levels to avoid low engagement, but whether levels have such effects in MI-BCI training still requires empirical validation. Support also lacks for the effectiveness of leaderboards that would allow social comparison [89,90,92]; narratives that contextualize the training and engage users [85,88,90]; personalization of the experience; and challenges that the user must overcome. The effectiveness of these elements has been supported by theoretical principles [65], hence their potential for MI-BCI training is worth examining in future studies.

Specifically, we propose a focus on rewards and goals, two other neglected elements that show promise for MI-BCI training. In contrast to other fields [88], rewards were not that common in MI-BCI training protocols. Furthermore, they were often presented in the form of points. Badges, frequently used in gamification of other activities [35,67], were not used in MI-BCI training. These forms of rewards have the potential to positively reinforce in-game actions mapped to specific MI tasks [211], which might subsequently improve BCI performance. In addition, users can be penalized for unsuccessful actions to discourage them from performing the MI task incorrectly. However, it is yet to be determined whether penalties are beneficial, as giving only positive feedback could improve user performance compared to giving both positive and negative feedback [44,137]. Future experimental investigations into rewards and penalties are needed to reveal effective strategies for reinforcing MI-BCI skill learning.

In contrast to rewards, goals were very common in MI-BCI training, varying in scale and content. Notable was the frequent use of object-oriented goals, where users performed actions in order to interact with an object in the virtual environment. Aside from functioning as implicit guidance, object-directed visual stimulation is known to help with motor imagery and offline BCI performance [212–214]. However, the role of objects during real-time MI-BCI control remains unstudied. Mladenović [22] suggested several object properties derived from human-computer interaction that could assist BCI training. For instance, an object's appearance can imply possible ways of interacting with it, hence 'affording' actions [196–198]. Such perceptual affordances have been extensively studied in the field of human-computer interaction [215,216], and neuroscientific research shows their ability to facilitate motor imagery [217]. Therefore, objects and their properties, such as the three-dimensionality of virtual objects, are a promising direction for future BCI research.

Our final recommendation for gamification is to account for a user's profile, which has shown to be of significance to training outcomes [13,25,26,30,168,218]. Specifically, several reviewed studies alluded to differential effects of game elements depending on the user's BCI skill. For instance, assistance can help users with low BCI control perceive higher control, while it obstructs users with high BCI control [155]. Conversely, high-performing users benefit more from collaboration, while it makes no difference for low-performing users whether they play alone or with others [145]. As agency correlated with performance for discrete feedback, but not continuous [158], the latter can be used to still give low-performing users a high sense of control over virtual limbs. Lastly, tactile feedback in the form of FES can help improve the accuracy of users with low control, alleviating BCI inefficiency [178]. Altogether, these findings suggest that the effects of game elements on MI-BCI training depend on user traits, providing another direction for future research. Based on the hitherto discussed findings, Table 4 summarizes our current and future recommendations for gamifying MI-BCI training.

### 5.2. Limitations and Future Recommendations

The observed positive impacts of gamification are promising, but there is still a need for more investigation into the isolated effects of individual elements. Although this review only reported effectiveness for elements that were validated with a control or comparison condition, not all findings constitute strong evidence due to small sample sizes and lack of proper statistical testing. More empirical control and rigorous statistical testing are required in future research to allow meta-analysis of the individual effects of game elements, elucidating not only whether gamification works, but also to what extent each game element impacts MI-BCI training outcomes. How elements interact is another avenue for future research, as often multiple elements are implemented together and their effects on user motivation are not always additive [76,219].

Table 4

*List of recommendations for effective gamification of MI-BCI training protocols. Recommendations marked with ○ require empirical support from future studies.*

| Element | Recommendation |
|---|---|
| Feedback | • Immerse the user into a three-dimensional environment with virtual reality<br>• Give visual feedback through interactive objects and avatars<br>• Provide haptic feedback through vibrations or electro-stimulation<br>• Tailor the amount of positive and negative feedback to the user |
| Avatar | • Embody the user in a human avatar that is controlled with congruent motor imagery<br>• Show the avatar from a first-person perspective with human arms visible |
| Assistance | • Adapt the difficulty level to the user's skill level<br>• Provide specific instructions about the movements to imagine<br>• Allow users to execute movements before motor imagery to help recall the kinesthetic sensation<br>• Avoid over-assisting users as this can frustrate and reduce their agency<br>○ Explain to the user how to use the BCI and play the game through an integrated tutorial |
| Social interaction | • Implement a multi-user system where users collaborate or compete in real-time<br>• Allow users to choose between single-player and multiplayer as this can help low-performers |
| Progress | ○ Indicate how far along the training the user has progressed through a percentage or bar |
| Score | ○ Quantify user performance across trials as a numeric game score |
| Timer | ○ Communicate how much time is left or spent to pressure the user |
| Goals | ○ Set short- and long-term goals to guide users and give their actions purpose, ideally at each time scale: per trial, per session, and for the whole experiment<br>○ Present interactive objects that afford actions and implicitly cue necessary motor imagery |
| Rewards | ○ Reward successful motor imagery execution with points or other incentives matching the game theme to positively reinforce desired user behavior and intrinsically motivate subjects<br>○ Punish unsuccessful motor imagery attempts with penalties such as losing points or progress |
| Challenges | ○ When the BCI can detect the strength of intended movements at any time the user desires (asynchronous BCI), challenge users behaviorally by having them perform motor imagery with specific strength, duration, or timing.<br>○ When the BCI only allows discrete commands at fixed intervals (synchronous BCI), challenge users cognitively by having them solve puzzles or think strategically |

Furthermore, studies should share and report their protocols in more detail with rationale for design choices, especially since most BCI games are developed from scratch by researchers [10,39]. If need be, this can be in a separate design document as supplementary materials rather than the main manuscript [220]. For this reason, it is also important that the field agrees on a definition and taxonomy for gamification that can be utilized for reporting. We hope that this systematic review sets the foundation of such a framework for the implementation and reporting of gamified MI-BCI training protocols. As mentioned before, this review included game elements common in most taxonomies, but some important elements could still have been overlooked. It should also be acknowledged that negative and null results might be underreported due to publication bias [221,222]. Therefore, there is a need for more rigorous and transparent reporting.

The positive results observed in this review mostly pertain to short-term effects, where the training took place over a few sessions. Longitudinal studies examining the long-term effects of gamified MI-BCI training are warranted, especially for use in physical rehabilitation which spans months to years. Finally, this review focused solely on the protocols that were used during real-time interaction with MI-BCI control. Some studies used a different protocol for calibration, such as the standard Graz protocol, which could have influenced BCI performance because the system was first trained with data collected in a non-gamified calibration session. In addition, we could not identify whether classifiers were updated after each session when new data became available. This makes it difficult to determine whether user performance increased due to the system learning from the user or the user learning from the gamified protocol.

## 5.3. Conclusion

This systematic review explored the state of the art of gamification in MI-BCI training by reviewing the protocols of 86 studies from the past decade. Overall, these studies point to the positive impacts of gamification in MI-BCI training, improving the performance and experience of users when learning BCI control using their motor imagery. We identified 14 game elements that had been implemented in the included records, although their prevalence across studies differed greatly. Our findings suggest that the use of specific game elements, such as avatars and assistance, can induce more agency and ownership over virtual limbs and physical robots, reduce the mental workload of users, motivate users, and make the overall training process more fun. Although promising, more empirical investigation is warranted into the effects of individual game elements to develop a full picture of how gamification can best help MI-BCI training. This is especially important for efficient MI-BCI use in real-world settings, such as neurorehabilitation programs. The numerous recommendations made in this review set the first step towards a framework for more effective, user-centered MI-BCI training, which ultimately may help alleviate BCI inefficiency.

## References

[1] Wolpaw J R, Millán J del R and Ramsey N F 2020 Chapter 2 - Brain-computer interfaces: Definitions and principles *Handbook of Clinical Neurology* Brain-Computer Interfaces vol 168, ed N F Ramsey and J del R Millán (Elsevier) pp 15–23

[2] Kawala-Sterniuk A, Browarska N, Al-Bakri A, Pelc M, Zygarlicki J, Sidikova M, Martinek R and Gorzelanczyk E J 2021 Summary of over Fifty Years with Brain-Computer Interfaces—A Review *Brain Sci* **11** 43

[3] Kübler A 2020 The history of BCI: From a vision for the future to real support for personhood in people with locked-in syndrome *Neuroethics* **13** 163–80

[4] Rezeika A, Benda M, Stawicki P, Gembler F, Saboor A and Volosyak I 2018 Brain–Computer Interface Spellers: A Review *Brain Sciences* **8** 57

[5] He Y, Eguren D, Azorín J M, Grossman R G, Luu T P and Contreras-Vidal J L 2018 Brain–machine interfaces for controlling lower-limb powered robotic systems *J. Neural Eng.* **15** 021004

[6] Bamdad M, Zarshenas H and Auais M A 2015 Application of BCI systems in neurorehabilitation: a scoping review *Disability and Rehabilitation: Assistive Technology* **10** 355–64

[7] Saha S, Mamun K A, Ahmed K, Mostafa R, Naik G R, Darvishi S, Khandoker A H and Baumert M 2021 Progress in Brain Computer Interface: Challenges and Opportunities *Frontiers in Systems Neuroscience* **15**

[8] Abdulkader S N, Atia A and Mostafa M-S M 2015 Brain computer interfacing: Applications and challenges *Egyptian Informatics Journal* **16** 213–30

[9] Wadeson A, Nijholt A and Nam C S 2015 Artistic brain-computer interfaces: state-of-the-art control mechanisms *Brain-Computer Interfaces* **2** 70–5

[10] Ahn M, Lee M, Choi J and Jun S C 2014 A Review of Brain-Computer Interface Games and an Opinion Survey from Researchers, Developers and Users *Sensors* **14** 14601–33

[11] Spüler M, Krumpe T, Walter C, Scharinger C, Rosenstiel W and Gerjets P 2017 Brain-Computer Interfaces for Educational Applications *Informational Environments : Effects of Use, Effective Designs* ed J Buder and F W Hesse (Cham: Springer International Publishing) pp 177–201

[12] Yadava M, Kumar P, Saini R, Roy P P and Prosad Dogra D 2017 Analysis of EEG signals and its application to neuromarketing *Multimed Tools Appl* **76** 19087–111

[13] Jeunet C, N'Kaoua B and Lotte F 2016 Chapter 1 - Advances in user-training for mental-imagery-based BCI control: Psychological and cognitive factors and their neural correlates *Progress in Brain Research* Brain-Computer Interfaces: Lab Experiments to Real-World Applications vol 228, ed D Coyle (Elsevier) pp 3–35

[14] Jeunet C, Jahanpour E and Lotte F 2016 Why standard brain-computer interface (BCI) training protocols should be changed: an experimental study *J. Neural Eng.* **13** 036024

[15] Zhang R, Li F, Zhang T, Yao D and Xu P 2020 Subject inefficiency phenomenon of motor imagery brain-computer interface: Influence factors and potential solutions *Brain Science Advances* **6** 224–41

[16] Allison B Z and Neuper C 2010 Could Anyone Use a BCI? *Brain-Computer Interfaces: Applying our Minds to Human-Computer Interaction* Human-Computer Interaction Series ed D S Tan and A Nijholt (London: Springer) pp 35–54

[17] Thompson M C 2019 Critiquing the Concept of BCI Illiteracy *Sci Eng Ethics* **25** 1217–33

[18] Lee S, Jang S and Jun S 2022 Exploring the Ability to Classify Visual Perception and Visual Imagery EEG Data: Toward an Intuitive BCI System *ELECTRONICS* **11**

[19] Kübler A, Holz E M, Riccio A, Zickler C, Kaufmann T, Kleih S C, Staiger-Sälzer P, Desideri L, Hoogerwerf E-J and Mattia D 2014 The User-Centered Design as Novel Perspective for Evaluating the Usability of BCI-Controlled Applications *PLOS ONE* **9** e112392

[20] Lécuyer A, Lotte F, Reilly R B, Leeb R, Hirose M and Slater M 2008 Brain-Computer Interfaces, Virtual Reality, and Videogames *Computer* **41** 66–72

[21] Mercado-García V R and Alonso-Valerdi L M 2018 The identity of the protagonist in brain-computer interfaces: A user-centered design approach *2018 International Conference on Electronics, Communications and Computers (CONIELECOMP)* 2018 International Conference on Electronics, Communications and Computers (CONIELECOMP) pp 44–9

[22] Mladenović J 2021 Standardization of protocol design for user training in EEG-based brain–computer interface *J. Neural Eng.* **18** 011003

[23] Plass-Oude Bos D, Poel M and Nijholt A 2011 A Study in User-Centered Design and Evaluation of Mental Tasks for BCI *Advances in Multimedia Modeling* Lecture Notes in Computer Science ed K-T Lee, W-H Tsai, H-Y M Liao, T Chen, J-W Hsieh and C-C Tseng (Berlin, Heidelberg: Springer) pp 122–34

[24] Roc A, Pillette L, Mladenovic J, Benaroch C, N'Kaoua B, Jeunet C and Lotte F 2021 A review of user training methods in brain computer interfaces based on mental tasks *J. Neural Eng.* **18** 011002

[25] Alimardani M and Gherman D-E 2022 Individual Differences in Motor Imagery BCIs: a Study of Gender, Mental States and Mu Suppression *2022 10th International Winter Conference on Brain-Computer Interface (BCI)* 2022 10th International Winter Conference on Brain-Computer Interface (BCI) pp 1–7

[26] Jeunet C, N'Kaoua B, Subramanian S, Hachet M and Lotte F 2015 Predicting Mental Imagery-Based BCI Performance from Personality, Cognitive Profile and Neurophysiological Patterns *PLOS ONE* **10** e0143962

[27] Kleih S C and Kübler A 2015 Psychological Factors Influencing Brain-Computer Interface (BCI) Performance *2015 IEEE International Conference on Systems, Man, and Cybernetics* 2015 IEEE International Conference on Systems, Man, and Cybernetics pp 3192–6

[28] Leeuwis N and Alimardani M 2020 High Aptitude Motor-Imagery BCI Users Have Better Visuospatial Memory *2020 IEEE International Conference on Systems, Man, and Cybernetics (SMC)* 2020 IEEE International Conference on Systems, Man, and Cybernetics (SMC) pp 1518–23

[29] Nijboer F, Birbaumer N and Kubler A 2010 The Influence of Psychological State and Motivation on Brain–Computer Interface Performance in Patients with Amyotrophic Lateral Sclerosis – a Longitudinal Study *Frontiers in Neuroscience* **4**

[30] Vourvopoulos A, Niforatos E, Hlinka M, Škola F and Liarokapis F 2017 Investigating the effect of user profile during training for BCI-based games *2017 9th International Conference on Virtual Worlds and Games for Serious Applications (VS-Games)* 2017 9th International Conference on Virtual Worlds and Games for Serious Applications (VS-Games) pp 117–24

[31] Lotte F, Faller J, Guger C, Renard Y, Pfurtscheller G, Lécuyer A and Leeb R 2013 Combining BCI with Virtual Reality: Towards New Applications and Improved BCI *Towards Practical Brain-Computer Interfaces: Bridging the Gap from Research to Real-World Applications* Biological and Medical Physics, Biomedical Engineering ed B Z Allison, S Dunne, R Leeb, J Del R. Millán and A Nijholt (Berlin, Heidelberg: Springer) pp 197–220

[32] Plass-Oude Bos D, Reuderink B, van de Laar B, Gürkök H, Mühl C, Poel M, Nijholt A and Heylen D 2010 Brain-Computer Interfacing and Games *Brain-Computer Interfaces: Applying our Minds to Human-Computer Interaction* Human-Computer Interaction Series ed D S Tan and A Nijholt (London: Springer) pp 149–78

[33] Škola F, Tinková S and Liarokapis F 2019 Progressive Training for Motor Imagery Brain-Computer Interfaces Using Gamification and Virtual Reality Embodiment *Frontiers in Human Neuroscience* **13**

[34] Deterding S, Dixon D, Khaled R and Nacke L 2011 From game design elements to gamefulness: defining "gamification" *Proceedings of the 15th International Academic MindTrek Conference: Envisioning Future Media Environments* MindTrek '11 (New York, NY, USA: Association for Computing Machinery) pp 9–15

[35] Manzano-León A, Camacho-Lazarraga P, Guerrero M A, Guerrero-Puerta L, Aguilar-Parra J M, Trigueros R and Alias A 2021 Between Level Up and Game Over: A Systematic Literature Review of Gamification in Education *Sustainability* **13** 2247

[36] Saleem A N, Noori N M and Ozdamli F 2022 Gamification Applications in E-learning: A Literature Review *Tech Know Learn* **27** 139–59

[37] Johnson D, Deterding S, Kuhn K-A, Staneva A, Stoyanov S and Hides L 2016 Gamification for health and wellbeing: A systematic review of the literature *Internet Interventions* **6** 89–106

[38] Larson K 2020 Serious Games and Gamification in the Corporate Training Environment: a Literature Review *TechTrends* **64** 319–28

[39] Vasiljevic G A M and de Miranda L C 2020 Brain–Computer Interface Games Based on Consumer-Grade EEG Devices: A Systematic Literature Review *International Journal of Human–Computer Interaction* **36** 105–42

[40] Gao C, Xia M, Zhang Z, Han Y and Gu Y 2023 Improving the Brain-Computer Interface Learning Process with Gamification in Motor Imagery: A Review *Gamification - Analysis, Design, Development and Ludification* ed I Deliyannis (IntechOpen)

[41] Lotte F, Bougrain L, Cichocki A, Clerc M, Congedo M, Rakotomamonjy A and Yger F 2018 A review of classification algorithms for EEG-based brain–computer interfaces: a 10 year update *J. Neural Eng.* **15** 031005

[42] Nicolas-Alonso L F and Gomez-Gil J 2012 Brain Computer Interfaces, a Review *Sensors* **12** 1211–79

[43] Wolpaw J R, Birbaumer N, McFarland D J, Pfurtscheller G and Vaughan T M 2002 Brain–computer interfaces for communication and control *Clinical Neurophysiology* **113** 767–91

[44] Alimardani M, Nishio S and Ishiguro H 2015 BCI-teleoperated androids; a study of embodiment and its effect on motor imagery learning *2015 IEEE 19th International Conference on Intelligent Engineering Systems (INES)* 2015 IEEE 19th International Conference on Intelligent Engineering Systems (INES) pp 347–52

[45] Vourvopoulos A, Ferreira A and Badia S B I 2016 NeuRow: An immersive VR environment for motor-imagery training with the use of Brain-Computer Interfaces and vibrotactile feedback *PhyCS 2016 - Proceedings of the 3rd International Conference on Physiological Computing Systems* pp 43–53

[46] Zander T O, Kothe C, Jatzev S and Gaertner M 2010 Enhancing Human-Computer Interaction with Input from Active and Passive Brain-Computer Interfaces *Brain-Computer Interfaces: Applying our Minds to Human-Computer Interaction* Human-Computer Interaction Series ed D S Tan and A Nijholt (London: Springer) pp 181–99

[47] Alimardani M and Hiraki K 2020 Passive Brain-Computer Interfaces for Enhanced Human-Robot Interaction *Frontiers in Robotics and AI* **7**

[48] Atilla F and Alimardani M 2021 EEG-based Classification of Drivers Attention using Convolutional Neural Network *2021 IEEE 2nd International Conference on Human-Machine Systems (ICHMS)* 2021 IEEE 2nd International Conference on Human-Machine Systems (ICHMS) pp 1–4

[49] Abiri R, Borhani S, Sellers E W, Jiang Y and Zhao X 2019 A comprehensive review of EEG-based brain–computer interface paradigms *J. Neural Eng.* **16** 011001

[50] Neuper C, Wörtz M and Pfurtscheller G 2006 ERD/ERS patterns reflecting sensorimotor activation and deactivation *Progress in Brain Research* Event-Related Dynamics of Brain Oscillations vol 159, ed C Neuper and W Klimesch (Elsevier) pp 211–22

[51] Neuper C, Scherer R, Wriessnegger S and Pfurtscheller G 2009 Motor imagery and action observation: Modulation of sensorimotor brain rhythms during mental control of a brain–computer interface *Clinical Neurophysiology* **120** 239–47

[52] Pfurtscheller G and Lopes da Silva F H 1999 Event-related EEG/MEG synchronization and desynchronization: basic principles *Clinical Neurophysiology* **110** 1842–57

[53] Jeannerod M 1995 Mental imagery in the motor context *Neuropsychologia* **33** 1419–32

[54] Jeon Y, Nam C S, Kim Y-J and Whang M C 2011 Event-related (De)synchronization (ERD/ERS) during motor imagery tasks: Implications for brain–computer interfaces *International Journal of Industrial Ergonomics* **41** 428–36

[55] Irimia D C, Ortner R, Poboroniuc M S, Ignat B E and Guger C 2018 High Classification Accuracy of a Motor Imagery Based Brain-Computer Interface for Stroke Rehabilitation Training *Frontiers in Robotics and AI* **5**

[56] Vavoulis A, Figueiredo P and Vourvopoulos A 2023 A Review of Online Classification Performance in Motor Imagery-Based Brain–Computer Interfaces for Stroke Neurorehabilitation *Signals* **4** 73–86

[57] Aggarwal S and Chugh N 2019 Signal processing techniques for motor imagery brain computer interface: A review *Array* **1–2** 100003

[58] Jaipriya D and Sriharipriya K C 2023 Brain Computer Interface-Based Signal Processing Techniques for Feature Extraction and Classification of Motor Imagery Using EEG: A Literature Review *Biomedical Materials & Devices*

[59] Pfurtscheller G and Neuper C 2001 Motor imagery and direct brain-computer communication *Proceedings of the IEEE* **89** 1123–34

[60] Vidaurre C and Blankertz B 2010 Towards a Cure for BCI Illiteracy *Brain Topogr* **23** 194–8

[61] Hammer E M, Halder S, Blankertz B, Sannelli C, Dickhaus T, Kleih S, Müller K-R and Kübler A 2012 Psychological predictors of SMR-BCI performance *Biological Psychology* **89** 80–6

[62] Horowitz A, Guger C and Korostenskaja M 2021 What Internal Variables Affect Sensorimotor Rhythm Brain-Computer Interface (SMR-BCI) Performance? *HCA Healthcare Journal of Medicine* **2**

[63] Pagulayan R J, Keeker K, Fuller T, Wixon D and Romero R L 2012 User-Centered Design in Games *Human Computer Interaction Handbook* (CRC Press)

[64] Mora A, Riera D, González C and Arnedo-Moreno J 2017 Gamification: a systematic review of design frameworks *J Comput High Educ* **29** 516–48

[65] Krath J, Schürmann L and von Korflesch H F O 2021 Revealing the theoretical basis of gamification: A systematic review and analysis of theory in research on gamification, serious games and game-based learning *Computers in Human Behavior* **125** 106963

[66] Buckley J, DeWille T, Exton C, Exton G and Murray L 2018 A Gamification–Motivation Design Framework for Educational Software Developers *Journal of Educational Technology Systems* **47** 101–27

[67] Hamari J, Koivisto J and Sarsa H 2014 Does Gamification Work? -- A Literature Review of Empirical Studies on Gamification *2014 47th Hawaii International Conference on System Sciences* 2014 47th Hawaii International Conference on System Sciences (HICSS) (Waikoloa, HI: IEEE) pp 3025–34

[68] Jia Y, Xu B, Karanam Y and Voida S 2016 Personality-targeted Gamification: A Survey Study on Personality Traits and Motivational Affordances *Proceedings of the 2016 CHI Conference on Human Factors in Computing Systems* CHI ’16 (New York, NY, USA: Association for Computing Machinery) pp 2001–13

[69] Weiser P, Bucher D, Cellina F and Luca V D 2015 A Taxonomy of Motivational Affordances for Meaningful Gamified and Persuasive Technologies EnviroInfo and ICT for Sustainability 2015 (Atlantis Press) pp 271–80

[70] Buckley P, Noonan S, Geary C, Mackessy T and Nagle E 2019 An Empirical Study of Gamification Frameworks *JOEUC* **31** 22–38

[71] Dicheva D, Dichev C, Agre G and Angelova G 2015 Gamification in Education: A Systematic Mapping Study *Journal of Educational Technology & Society* **18** 75–88

[72] Bedwell W L, Pavlas D, Heyne K, Lazzara E H and Salas E 2012 Toward a Taxonomy Linking Game Attributes to Learning: An Empirical Study *Simulation & Gaming* **43** 729–60

[73] Seaborn K and Fels D I 2015 Gamification in theory and action: A survey *International Journal of Human-Computer Studies* **74** 14–31

[74] Chou Y 2019 *Actionable Gamification: Beyond Points, Badges, and Leaderboards* (Packt Publishing Ltd)

[75] Aparicio A F, Vela F L G, Sánchez J L G and Montes J L I 2012 Analysis and application of gamification *Proceedings of the 13th International Conference on Interacción Persona-Ordenador* INTERACCION '12 (New York, NY, USA: Association for Computing Machinery) pp 1–2

[76] Dichev C and Dicheva D 2017 Gamifying education: what is known, what is believed and what remains uncertain: a critical review *International Journal of Educational Technology in Higher Education* **14** 9

[77] Werbach K 2014 (Re)Defining Gamification: A Process Approach *Persuasive Technology* Lecture Notes in Computer Science ed A Spagnolli, L Chittaro and L Gamberini (Cham: Springer International Publishing) pp 266–72

[78] Matallaoui A, Koivisto J, Hamari J and Zarnekow R 2017 *How Effective Is "Exergamification"? A Systematic Review on the Effectiveness of Gamification Features in Exergames*

[79] Xu F, Buhalis D and Weber J 2017 Serious games and the gamification of tourism *Tourism Management* **60** 244–56

[80] Wanick V and Bui H 2019 Gamification in Management: a systematic review and research directions *International Journal of Serious Games* **6** 57–74

[81] Brown M, O'Neill N, Woerden H van, Eslambolchilar P, Jones M and John A 2016 Gamification and Adherence to Web-Based Mental Health Interventions: A Systematic Review *JMIR Mental Health* **3** e5710

[82] Fleming T M, Bavin L, Stasiak K, Hermansson-Webb E, Merry S N, Cheek C, Lucassen M, Lau H M, Pollmuller B and Hetrick S 2017 Serious Games and Gamification for Mental Health: Current Status and Promising Directions *Frontiers in Psychiatry* **7**

[83] Mubin O, Alnajjar F, Jishtu N, Alsinglawi B and Mahmud A A 2019 Exoskeletons With Virtual Reality, Augmented Reality, and Gamification for Stroke Patients' Rehabilitation: Systematic Review *JMIR Rehabilitation and Assistive Technologies* **6** e12010

[84] Sardi L, Idri A and Fernández-Alemán J L 2017 A systematic review of gamification in e-Health *Journal of Biomedical Informatics* **71** 31–48

[85] Ferro L S 2021 The Game Element and Mechanic (GEM) framework: A structural approach for implementing game elements and mechanics into game experiences *Entertainment Computing* **36** 100375

[86] Radianti J, Majchrzak T A, Fromm J and Wohlgenannt I 2020 A systematic review of immersive virtual reality applications for higher education: Design elements, lessons learned, and research agenda *Computers & Education* **147** 103778

[87] Tomé Klock A C, da Cunha L F, de Carvalho M F, Eduardo Rosa B, Jaqueline Anton A and Gasparini I 2015 Gamification in e-Learning Systems: A Conceptual Model to Engage Students and Its Application in an Adaptive e-Learning System *Learning and Collaboration Technologies* Lecture Notes in Computer Science ed P Zaphiris and A Ioannou (Cham: Springer International Publishing) pp 595–607

[88] Vermeir J F, White M J, Johnson D, Crombez G and Ryckeghem D M L V 2020 The Effects of Gamification on Computerized Cognitive Training: Systematic Review and Meta-Analysis *JMIR Serious Games* **8** e18644

[89] Toda A M, Klock A C T, Oliveira W, Palomino P T, Rodrigues L, Shi L, Bittencourt I, Gasparini I, Isotani S and Cristea A I 2019 Analysing gamification elements in educational environments using an existing Gamification taxonomy *Smart Learn. Environ.* **6** 16

[90] Sailer M, Hense J U, Mayr S K and Mandl H 2017 How gamification motivates: An experimental study of the effects of specific game design elements on psychological need satisfaction *Computers in Human Behavior* **69** 371–80

[91] Tuah N M, Ahmedy F, Gani A and Yong L N 2021 A Survey on Gamification for Health Rehabilitation Care: Applications, Opportunities, and Open Challenges *Information* **12** 91

[92] Thiebes S, Lins S and Basten D 2014 Gamifying Information Systems - a synthesis of Gamification mechanics and Dynamics Twenty Second European Conference on Information Systems (Tel Aviv)

[93] Tondello G F, Mora A and Nacke L E 2017 Elements of Gameful Design Emerging from User Preferences *Proceedings of the Annual Symposium on Computer-Human Interaction in Play* CHI PLAY '17 (New York, NY, USA: Association for Computing Machinery) pp 129–42

[94] Ahmad M 2019 Categorizing Game Design Elements into Educational Game Design Fundamentals *Game Design and Intelligent Interaction* (IntechOpen)

[95] Ferreira A L S, Marciano J N, Miranda L C de and Miranda E E C de 2014 Understanding and Proposing a Design Rationale of Digital Games based on Brain-Computer Interface: Results of the AdmiralMind Battleship Study *Journal on Interactive Systems* **5**

[96] Lecuyer A, Lotte F, Reilly R B, Leeb R, Hirose M and Slater M 2008 Brain-Computer Interfaces, Virtual Reality, and Videogames *Computer* **41** 66–72

[97] Lotte F 2011 Brain-computer interfaces for 3D games: hype or hope? *Proceedings of the 6th International Conference on Foundations of Digital Games* FDG '11 (New York, NY, USA: Association for Computing Machinery) pp 325–7

[98] Marshall D, Coyle D, Wilson S and Callaghan M 2013 Games, Gameplay, and BCI: The State of the Art *IEEE Trans. Comput. Intell. AI Games* **5** 82–99

[99] Murdoch R 2019 An Experiential Learning-Based Approach to Neurofeedback Visualisation in Serious Games *Biomedical Visualisation : Volume 3* Advances in Experimental Medicine and Biology ed P M Rea (Cham: Springer International Publishing) pp 97–109

[100] Scott J A and Sims M 2021 Acceleration of Therapeutic Use of Brain Computer Interfaces by Development for Gaming *Intelligent Technologies for Interactive Entertainment* Lecture Notes of the Institute for Computer Sciences, Social Informatics and Telecommunications Engineering ed N Shaghaghi, F Lamberti, B Beams, R Shariatmadari and A Amer (Cham: Springer International Publishing) pp 267–81

[101] Hadjiaros M, Neokleous K, Shimi A, Avraamides M N and Pattichis C S 2023 Virtual Reality Cognitive Gaming Based on Brain Computer Interfacing: A Narrative Review *IEEE Access* **11** 18399–416

[102] Cattan G, Andreev A and Visinoni E 2020 Recommendations for Integrating a P300-Based Brain–Computer Interface in Virtual Reality Environments for Gaming: An Update *Computers* **9** 92

[103] Al-Qaysi Z T, Ahmed M A, Hammash N M, Hussein A F, Albahri A S, Suzani M S and Al-Bander B 2022 A systematic rank of smart training environment applications with motor imagery brain-computer interface *Multimedia Tools and Applications*

[104] Badia S B i, Fluet G G, Llorens R and Deutsch J E 2016 Virtual Reality for Sensorimotor Rehabilitation Post Stroke: Design Principles and Evidence *Neurorehabilitation Technology* ed D J Reinkensmeyer and V Dietz (Cham: Springer International Publishing) pp 573–603

[105] Leeb R and Pérez-Marcos D 2020 Chapter 14 - Brain-computer interfaces and virtual reality for neurorehabilitation *Handbook of Clinical Neurology* Brain-Computer Interfaces vol 168, ed N F Ramsey and J del R Millán (Elsevier) pp 183–97

[106] Wang K, Tian F, Pan L, Xu M, Li M, Dong B and Ming D 2024 Virtual Reality Game-Based Adaptive Neurofeedback Training for Motor Imagery *12th Asian-Pacific Conference on Medical and Biological Engineering* IFMBE Proceedings vol 103, ed G Wang, D Yao, Z Gu, Y Peng, S Tong and C Liu (Cham: Springer Nature Switzerland) pp 296–303

[107] Scherer R, Müller-Putz G, Friedrich E V C, Pammer-Schindler V, Wilding K, Keller S and Pirker J 2017 Games for BCI Skill Learning *Handbook of Digital Games and Entertainment Technologies* ed R Nakatsu, M Rauterberg and P Ciancarini (Singapore: Springer Singapore) pp 173–96

[108] Page M J, McKenzie J E, Bossuyt P M, Boutron I, Hoffmann T C, Mulrow C D, Shamseer L, Tetzlaff J M, Akl E A, Brennan S E, Chou R, Glanville J, Grimshaw J M, Hróbjartsson A, Lalu M M, Li T, Loder E W, Mayo-Wilson E, McDonald S, McGuinness L A, Stewart L A, Thomas J, Tricco A C, Welch V A, Whiting P and Moher D 2021 The PRISMA 2020 statement: An updated guideline for reporting systematic reviews *International Journal of Surgery* **88** 105906

[109] Ortner R, Ram D, Kollreider A, Pitsch H, Wojtowicz J and Edlinger G 2013 *Human-computer confluence for rehabilitation purposes after stroke* vol 8022 LNCS

[110] de Castro-Cros M, Sebastian-Romagosa M, Rodriguez-Serrano J, Opisso E, Ochoa M, Ortner R, Guger C and Tost D 2020 Effects of Gamification in BCI Functional Rehabilitation *FRONTIERS IN NEUROSCIENCE* **14**

[111] Li T, Zhang J, Xue T and Wang B 2017 Development of a Novel Motor Imagery Control Technique and Application in a Gaming Environment *Computational Intelligence and Neuroscience : CIN* **2017**

[112] Hougaard B I, Knoche H, Kristensen M S and Jochumsen M 2022 Modulating Frustration and Agency Using Fabricated Input for Motor Imagery BCIs in Stroke Rehabilitation *IEEE Access* **10** 72312–27

[113] Achanccaray D, Pacheco K, Carranza E and Hayashibe M 2018 Immersive Virtual Reality Feedback in a Brain Computer Interface for Upper Limb Rehabilitation *2018 IEEE International Conference on Systems, Man, and Cybernetics (SMC)* 2018 IEEE International Conference on Systems, Man, and Cybernetics (SMC) pp 1006–10

[114] Arpaia P, Esposito A, Mancino F, Moccaldi N and Natalizio A 2021 *Active and Passive Brain-Computer Interfaces Integrated with Extended Reality for Applications in Health 4.0* vol 12980 LNCS

[115] Badia S B i, Morgade A G, Samaha H and Verschure P F M J 2013 Using a Hybrid Brain Computer Interface and Virtual Reality System to Monitor and Promote Cortical Reorganization through Motor Activity and Motor Imagery Training *IEEE Transactions on Neural Systems and Rehabilitation Engineering* **21** 174–81

[116] Bordoloi S, Sharmah U and Hazarika S M 2012 Motor imagery based BCI for a maze game *2012 4th International Conference on Intelligent Human Computer Interaction (IHCI)* 2012 4th International Conference on Intelligent Human Computer Interaction (IHCI) pp 1–6

[117] Cunha J D, Perdikis S, Halder S and Scherer R 2021 Post-Adaptation Effects in a Motor Imagery Brain-Computer Interface Online Coadaptive Paradigm *IEEE Access* **9** 41688–703

[118] Huang D, Qian K, Fei D-Y, Jia W, Chen X and Bai O 2012 Electroencephalography (EEG)-Based Brain–Computer Interface (BCI): A 2-D Virtual Wheelchair Control Based on Event-Related Desynchronization/Synchronization and State Control *IEEE Transactions on Neural Systems and Rehabilitation Engineering* **20** 379–88

[119] King C E, Wang P T, Chui L A, Do A H and Nenadic Z 2013 Operation of a brain-computer interface walking simulator for individuals with spinal cord injury. *Journal of neuroengineering and rehabilitation* **10** 77

[120] Kreilinger A, Hiebel H and Müller-Putz G R 2016 Single Versus Multiple Events Error Potential Detection in a BCI-Controlled Car Game With Continuous and Discrete Feedback *IEEE Transactions on Biomedical Engineering* **63** 519–29

[121] McMahon M and Schukat M 2018 A low-Cost, Open-Source, BCI- VR Game Control Development Environment Prototype for Game Based Neurorehabilitation *2018 IEEE Games, Entertainment, Media Conference (GEM)* 2018 IEEE Games, Entertainment, Media Conference (GEM) pp 1–9

[122] Ponferrada E G, Sylaidi A and Aldo Faisal A 2018 Data-efficient motor imagery decoding in real-time for the cybathlon brain-computer interface race *NEUROTECHNIX 2018 - Proceedings of the 6th International Congress on Neurotechnology, Electronics and Informatics* pp 21–32

[123] Prapas G, Glavas K, Tzallas A T, Tzimourta K D, Giannakeas N and Tsipouras M G 2022 Motor Imagery Approach for BCI Game Development *2022 7th South-East Europe Design Automation, Computer Engineering, Computer Networks and Social Media Conference (SEEDA-CECNSM)* 2022 7th South-East Europe Design Automation, Computer Engineering, Computer Networks and Social Media Conference (SEEDA-CECNSM) pp 1–5

[124] Saeedi S, Chavarriaga R, Leeb R and Millan J D R 2016 Adaptive Assistance for Brain-Computer Interfaces by Online Prediction of Command Reliability *IEEE Computational Intelligence Magazine* **11** 32–9

[125] Wang Z, Yang Y, Xu M, Liu Y, Yin E and Zhou Z 2019 Towards a Hybrid BCI Gaming Paradigm Based on Motor Imagery and SSVEP *International Journal of Human - Computer Interaction* **35** 197–205

[126] Xu J, Liu T-L, Wu Z, Wu Z, Li Y and Nürnberger A 2020 Neurorehabilitation System in Virtual Reality with Low-Cost BCI Devices *2020 IEEE International Conference on Human-Machine Systems (ICHMS)* 2020 IEEE International Conference on Human-Machine Systems (ICHMS) pp 1–3

[127] Morone G, Pisotta I, Pichiorri F, Kleih S, Paolucci S, Molinari M, Cincotti F, Kübler A and Mattia D 2015 Proof of principle of a brain-computer interface approach to support poststroke arm rehabilitation in hospitalized patients: Design, acceptability, and usability *Archives of Physical Medicine and Rehabilitation* **96** S71–8

[128] Tezza D, Garcia S, Hossain T and Andujar M 2019 *Brain eRacing: An Exploratory Study on Virtual Brain-Controlled Drones* vol 11575 LNCS

[129] Tezza D, Caprio D, Garcia S, Pinto B, Laesker D and Andujar M 2020 *Brain-controlled drone racing game: A qualitative analysis* vol 12211 LNCS

[130] Karácsony T, Hansen J P, Iversen H K and Puthusserypady S 2019 Brain computer interface for neuro-rehabilitation with deep learning classification and virtual reality feedback *ACM International Conference Proceeding Series*

[131] Liu Z, Bryan J, Borkoski R, Yuan F, Li Y and Zhao X 2020 On a gamified brain-computer interface for cognitive training of spatial working memory *ASME 2020 Dynamic Systems and Control Conference, DSCC 2020* vol 1

[132] Vourvopoulos A, Ferreira A and Badia S B i 2019 *Development and assessment of a self-paced BCI-VR paradigm using multimodal stimulation and adaptive performance* vol 10057 LNCS

[133] Taherian S, Selitskiy D, Pau J and Claire Davies T 2017 Are we there yet? Evaluating commercial grade brain-computer interface for control of computer applications by individuals with cerebral palsy. *Disability and rehabilitation. Assistive technology* **12** 165–74

[134] Achanccaray D, Mylonas G and Andreu-Perez J 2019 An Implicit Brain Computer Interface Supported by Gaze Monitoring for Virtual Therapy *2019 IEEE International Conference on Systems, Man and Cybernetics (SMC)* 2019 IEEE International Conference on Systems, Man and Cybernetics (SMC) pp 2829–32

[135] Achanccaray D, Izumi S-I and Hayashibe M 2021 Visual-Electrotactile Stimulation Feedback to Improve Immersive Brain-Computer Interface Based on Hand Motor Imagery *Computational Intelligence and Neuroscience : CIN* **2021**

[136] Alchalabi B and Faubert J 2019 A Comparison between BCI Simulation and Neurofeedback for Forward/Backward Navigation in Virtual Reality *Computational Intelligence and Neuroscience : CIN* **2019** 12

[137] Alchalabi B, Faubert J and Labbé D R 2021 A multi-modal modified feedback self-paced BCI to control the gait of an avatar. *Journal of neural engineering* **18**

[138] Arpaia P, Coyle D, Donnarumma F, Esposito A, Natalizio A, Parvis M, Pesola M and Vallefuoco E 2022 Multimodal Feedback in Assisting a Wearable Brain-Computer Interface Based on Motor Imagery *2022 IEEE International Conference on Metrology for Extended Reality, Artificial Intelligence and Neural Engineering (MetroXRAINE)* 2022 IEEE International Conference on Metrology for Extended Reality, Artificial Intelligence and Neural Engineering (MetroXRAINE) pp 691–6

[139] Arpaia P, D'Angelo M, D'Errico G, Paolis L T D, Esposito A, Grassini S, Moccaldi N, Natalizio A and Nuzzo B L 2022 EEG features of the interaction between sense of agency and body ownership: a motor imagery BCI case study

*2022 IEEE International Conference on Metrology for Extended Reality, Artificial Intelligence and Neural Engineering (MetroXRAINE)* 2022 IEEE International Conference on Metrology for Extended Reality, Artificial Intelligence and Neural Engineering (MetroXRAINE) pp 104–9

[140] Asensio-Cubero J, Gan J Q and Palaniappan R 2016 Multiresolution analysis over graphs for a motor imagery based online BCI game *Computers in Biology and Medicine* **68** 21–6

[141] Battison A, Schlussel M, Fuller T, Yu Y-C and Gabel L 2015 Effectiveness of subject specific instruction on mu-based brain-computer interface training *2015 41st Annual Northeast Biomedical Engineering Conference (NEBEC)* 2015 41st Annual Northeast Biomedical Engineering Conference (NEBEC) pp 1–2

[142] Bigirimana A D, Siddique N and Coyle D 2020 Emotion-Inducing Imagery Versus Motor Imagery for a Brain-Computer Interface *IEEE Transactions on Neural Systems and Rehabilitation Engineering* **28** 850–9

[143] Blanco-Mora D A, Aldridge A, Jorge C, Vourvopoulos A, Figueiredo P and Badia S B I 2022 Impact of age, VR, immersion, and spatial resolution on classifier performance for a MI-based BCI *Brain-Computer Interfaces* **9** 169–78

[144] Bois du, Bigirimana A, Korik A, Kéthina G, Rutembesa E, Mutabaruka J, Mutesa L, Prasad G, Jansen S and Coyle D 2021 Neurofeedback with low-cost, wearable electroencephalography (EEG) reduces symptoms in chronic post-traumatic stress disorder *Journal of Affective Disorders* **295** 1319–34

[145] Bonnet L, Lotte F and Lécuyer A 2013 Two Brains, One Game: Design and Evaluation of a Multiuser BCI Video Game Based on Motor Imagery *IEEE Trans. Comput. Intell. AI Games* **5** 185–98

[146] Cantillo-Negrete J, Carino-Escobar R I, Carrillo-Mora P, Barraza-Madrigal J A and Arias-Carrión O 2019 Robotic orthosis compared to virtual hand for Brain–Computer Interface feedback *Biocybernetics and Biomedical Engineering* **39** 263–72

[147] Cardoso V, Delisle-Rodriguez D, Romero-Laiseca M, Loterio F, Gurve D, Floriano A, Valadao C, Silva L, Krishnan S, Frizera-Neto A and Bastos T 2021 Effect of a Brain-Computer Interface Based on Pedaling Motor Imagery on Cortical Excitability and Connectivity *SENSORS* **21**

[148] Choi J, Huh S and Jo S 2020 Improving performance in motor imagery BCI-based control applications via virtually embodied feedback *COMPUTERS IN BIOLOGY AND MEDICINE* **127**

[149] Da Silva-Sauer L, Valero-Aguayo L, Velasco-Álvarez F, Varona-Moya S and Ron-Angevin R 2015 *Training in realistic virtual environments: Impact on user performance in a motor imagery-based brain–computer interface* vol 9094

[150] Davis D L, Nakanishi M and Jung T-P 2022 A Comparison Study of Egocentric and Allocentric Visual Feedback for Motor-Imagery Brain-Computer Interfaces *2022 IEEE International Conference on Systems, Man, and Cybernetics (SMC)* 2022 IEEE International Conference on Systems, Man, and Cybernetics (SMC) pp 1630–5

[151] Duan F, Lin D, Li W and Zhang Z 2015 Design of a Multimodal EEG-based Hybrid BCI System with Visual Servo Module *IEEE Transactions on Autonomous Mental Development* **7** 332–41

[152] Espinosa A, Ortner R, Irimia D and Guger C 2012 Rehabilitation through brain computer interfaces classification and feedback study *IJCCI 2012 - Proceedings of the 4th International Joint Conference on Computational Intelligence* pp 692–7

[153] Ferrero L, Ortiz M, Quiles V, Iáñez E and Azorín J M 2021 Improving Motor Imagery of Gait on a Brain–Computer Interface by Means of Virtual Reality: A Case of Study *IEEE Access* **9** 49121–30

[154] Hasan B and Gan J 2012 Hangman BCI: An unsupervised adaptive self-paced Brain-Computer Interface for playing games *COMPUTERS IN BIOLOGY AND MEDICINE* **42** 598–606

[155] Jochumsen M, Hougaard B I, Kristensen M S and Knoche H 2022 Implementing Performance Accommodation Mechanisms in Online BCI for Stroke Rehabilitation: A Study on Perceived Control and Frustration *Sensors* **22** 9051

[156] Juliano J, Spicer R, Vourvopoulos A, Lefebvre S, Jann K, Ard T, Santarnecchi E, Krum D and Liew S 2020 Embodiment Is Related to Better Performance on a Brain-Computer Interface in Immersive Virtual Reality: A Pilot Study *SENSORS* **20**

[157] Kapitonova M, Haeringer Z, Bongk E and Ball T 2022 A Modular, Adaptive, Deep-Learning-Based Brain-VR Interface *bioRxiv*

[158] Kjeldsen T K K, Nielsen T B, Ziadeh H, Lehmann S, Nielsen L D, Gulyás D, Hougaard B I, Knoche H and Jochumsen M 2021 Effect of Continuous and Discrete Feedback on Agency and Frustration in a Brain-Computer Interface Virtual Reality Interaction *2021 IEEE 21st International Conference on Bioinformatics and Bioengineering (BIBE)* 2021 IEEE 21st International Conference on Bioinformatics and Bioengineering (BIBE) pp 1–5

[159] Kosmyna N, Tarpin-Bernard F and Rivet B 2014 Bidirectional feedback in motor imagery BCIs: Learn to control a drone within 5 minutes *CHI'14 Extended Abstracts on Human Factors in Computing Systems* pp 479–82

[160] Kosmyna N, Tarpin-Bernard F and Rivet B 2015 Adding Human Learning in Brain-Computer Interfaces (BCIs): Towards a Practical Control Modality *ACM TRANSACTIONS ON COMPUTER-HUMAN INTERACTION* **22**

[161] Lechner A, Ortner R and Guger C 2014 *Feedback strategies for BCI based stroke rehabilitation: Evaluation of different approaches* vol 7

[162] Leeb R, Lancelle M, Kaiser V, Fellner D and Pfurtscheller G 2013 Thinking Penguin: Multimodal Brain-Computer Interface Control of a VR Game *IEEE TRANSACTIONS ON COMPUTATIONAL INTELLIGENCE AND AI IN GAMES* **5** 117–28

[163] Long J, Li Y, Wang H, Yu T, Pan J and Li F 2012 A Hybrid Brain Computer Interface to Control the Direction and Speed of a Simulated or Real Wheelchair *IEEE Transactions on Neural Systems and Rehabilitation Engineering* **20** 720–9

[164] Long J, Li Y, Wang H, Yu T and Pan J 2012 Control of a simulated wheelchair based on a hybrid brain computer interface *2012 Annual International Conference of the IEEE Engineering in Medicine and Biology Society* 2012 Annual International Conference of the IEEE Engineering in Medicine and Biology Society pp 6727–30

[165] Mahmood M, Kwon S, Kim H, Kim Y, Siriaraya P, Choi J, Otkhmezuri B, Kang K, Yu K J, Jang Y C, Ang C S and Yeo W 2021 Wireless Soft Scalp Electronics and Virtual Reality System for Motor Imagery-Based Brain–Machine Interfaces *Advanced Science* **8**

[166] McCreadie K A, Coyle D H and Prasad G 2014 Is Sensorimotor BCI Performance Influenced Differently by Mono, Stereo, or 3-D Auditory Feedback? *IEEE Transactions on Neural Systems and Rehabilitation Engineering* **22** 431–40

[167] Miao Y, Chen S, Zhang X, Jin J, Xu R, Daly I, Jia J, Wang X, Cichocki A and Jung T-P 2020 BCI-Based Rehabilitation on the Stroke in Sequela Stage *Neural Plasticity* **2020**

[168] Mladenović J, Frey J, Pramij S, Mattout J and Lotte F 2021 Towards Identifying Optimal Biased Feedback for Various User States and Traits in Motor Imagery BCI *IEEE Transactions on Biomedical Engineering* **69** 1101–10

[169] Mladenović J, Frey J, Bonnet-Save M, Mattout J and Lotte F 2017 The Impact of Flow in an EEG-based Brain Computer Interface *7th International BCI Conference*

[170] Myrden A and Chau T 2015 Effects of user mental state on EEG-BCI performance *FRONTIERS IN HUMAN NEUROSCIENCE* **9**

[171] Nicolelis M A L, Alho E J L, Donati A R C, Yonamine S, Aratanha M A, Bao G, Campos D S F, Almeida S, Fischer D and Shokur S 2022 Training with noninvasive brain–machine interface, tactile feedback, and locomotion to enhance neurological recovery in individuals with complete paraplegia: a randomized pilot study *Scientific Reports (Nature Publisher Group)* **12**

[172] Nierula B, Spanlang B, Martini M, Borrell M, Nikulin V V and Sanchez-Vives M V 2021 Agency and responsibility over virtual movements controlled through different paradigms of brain-computer interface. *The Journal of physiology* **599** 2419–34

[173] Novak D, Sigrist R, Gerig N, Wyss D, Bauer R, Gotz U and Riener R 2018 Benchmarking Brain-Computer Interfaces Outside the Laboratory:The Cybathlon 2016 *FRONTIERS IN NEUROSCIENCE* **11**

[174] Ortner R, Irimia D-C, Scharinger J and Guger C 2012 A Motor imagery based Brain-Computer Interface for stroke rehabilitation *Annual Review of CyberTherapy and Telemedicine* **10** 319–23

[175] Perdikis S, Tonin L, Saeedi S and Schneider C 2018 The Cybathlon BCI race: Successful longitudinal mutual learning with two tetraplegic users *PLoS Biology* **16**

[176] Petrushin A, Tessadori J, Barresi G and Mattos L S 2018 Effect of a Click-Like Feedback on Motor Imagery in EEG-BCI and Eye-Tracking Hybrid Control for Telepresence *2018 IEEE/ASME International Conference on Advanced Intelligent Mechatronics (AIM)* 2018 IEEE/ASME International Conference on Advanced Intelligent Mechatronics (AIM) pp 628–33

[177] Pradhapan P, Grossekathofer U, Schiavone G, Grundlehner B and Mihajlovic V 2018 *Toward Practical BCI Solutions for Entertainment and Art Performance* ed C Nam

[178] Ren S, Wang W, Hou Z-G, Liang X, Wang J and Shi W 2020 Enhanced Motor Imagery Based Brain- Computer Interface via FES and VR for Lower Limbs *IEEE Transactions on Neural Systems and Rehabilitation Engineering* **28** 1846–55

[179] Sakamaki I, Tavakoli M, Wiebe S and Adams K 2022 Examination of effectiveness of kinaesthetic haptic feedback for motor imagery-based brain-computer interface training *Brain-Computer Interfaces*

[180] Sebastián-Romagosa M, Cho W, Ortner R, Murovec N, Von Oertzen T, Kamada K, Allison B Z and Guger C 2020 Brain Computer Interface Treatment for Motor Rehabilitation of Upper Extremity of Stroke Patients—A Feasibility Study *Frontiers in Neuroscience*

[181] Shane N M, McCreadie D K, Charles D D, Korik D A and Coyle P D 2022 Online 3D Motion Decoder BCI for Embodied Virtual Reality Upper Limb Control: A Pilot Study *2022 IEEE International Conference on Metrology for Extended Reality, Artificial Intelligence and Neural Engineering (MetroXRAINE)* 2022 IEEE International Conference on Metrology for Extended Reality, Artificial Intelligence and Neural Engineering (MetroXRAINE) pp 697–702

[182] Škola F and Liarokapis F 2018 Embodied VR environment facilitates motor imagery brain–computer interface training *Computers and Graphics (Pergamon)* **75** 59–71

[183] Tabernig C B, Carrere L C, Manresa J B and Spaich E G 2021 Does feedback based on FES-evoked nociceptive withdrawal reflex condition event-related desynchronization? An exploratory study with brain-computer interfaces. *Biomedical physics & engineering express* **7**

[184] Tezza D, Caprio D, Pinto B, Mantilla I and Andujar M 2020 *An analysis of engagement levels while playing brain-controlled games* vol 12211 LNCS

[185] van de Laar B, Reuderink B, Bos D P-O and Heylen D 2012 Evaluating user experience of actual and imagined movement in BCI gaming *Interdisciplinary advancements in gaming, simulations, and virtual environments: Emerging trends.* ed R E Ferdig and S de Freitas (Hershey, PA: Information Science Reference/IGI Global) pp 266–80

[186] Vourvopoulos A and Badia S B i 2016 Motor priming in virtual reality can augment motor-imagery training efficacy in restorative brain-computer interaction: a within-subject analysis. *Journal of neuroengineering and rehabilitation* **13** 69

[187] Wang P T, King C E, Chui L A, Do A H and Nenadic Z 2012 Self-paced brain-computer interface control of ambulation in a virtual reality environment *Journal of Neural Engineering* **9**

[188] Xie P, Wang Z, Li Z, Wang Y, Wang N, Liang Z, Wang J and Chen X 2022 Research on Rehabilitation Training Strategies Using Multimodal Virtual Scene Stimulation *Frontiers in Aging Neuroscience* **14**

[189] Yang C, Ye Y, Li X and Wang R 2018 Development of a neuro-feedback game based on motor imagery EEG *Multimedia Tools and Applications* **77** 15929–49

[190] Ziadeh H, Gulyas D, Nielsen L D, Lehmann S, Nielsen T B, Kjeldsen T K K, Hougaard B I, Jochumsen M and Knoche H 2021 "Mine Works Better": Examining the Influence of Embodiment in Virtual Reality on the Sense of Agency During a Binary Motor Imagery Task With a Brain-Computer Interface. *Frontiers in psychology* **12** 806424

[191] Evans N, Gale S, Schurger A and Blanke O 2015 Visual Feedback Dominates the Sense of Agency for Brain-Machine Actions *PLOS ONE* **10** e0130019

[192] Thurlings M E, van Erp J B F, Brouwer A-M, Blankertz B and Werkhoven P 2012 Control-display mapping in brain–computer interfaces *Ergonomics* **55** 564–80

[193] Lemmens J S and von Münchhausen C F 2023 Let the beat flow: How game difficulty in virtual reality affects flow *Acta Psychologica* **232** 103812

[194] Yu D, Wang S, Song F, Liu Y, Zhang S, Wang Y, Xie X and Zhang Z 2023 Research on user experience of the video game difficulty based on flow theory and fNIRS *Behaviour & Information Technology* **42** 789–805

[195] Cao S and Liu F 2022 Learning to play: understanding in-game tutorials with a pilot study on implicit tutorials *Heliyon* **8** e11482

[196] Cardona-Rivera R E and Young R M 2013 A Cognitivist Theory of Affordances for Games *Proceedings of the 2013 DiGRA International Conference: DeFragging Game Studies* (Atlanta, GA)

[197] Pinchbeck D 2009 An affordance based model for gameplay *Proceedings of the 2009 DiGRA International Conference: Breaking New Ground: Innovation in Games, Play, Practice and Theory* (London, UK: Brunel University)

[198] Rambusch J and Susi T 2008 The Challenge of Managing Affordances in Computer Game Play *Human IT: Journal for Information Technology Studies as a Human Science* **9**

[199] Ryan R M, Rigby C S and Przybylski A 2006 The Motivational Pull of Video Games: A Self-Determination Theory Approach *Motiv Emot* **30** 344–60

[200] Tyack A and Wyeth P 2021 "The Small Decisions Are What Makes it Interesting": Autonomy, Control, and Restoration in Player Experience *Proc. ACM Hum.-Comput. Interact.* **5** 282:1-282:26

[201] Whyte E M, Smyth J M and Scherf K S 2015 Designing Serious Game Interventions for Individuals with Autism *J Autism Dev Disord* **45** 3820–31

[202] Krämer N, Kopp S, Becker-Asano C and Sommer N 2013 Smile and the world will smile with you—The effects of a virtual agent's smile on users' evaluation and behavior *International Journal of Human-Computer Studies* **71** 335–49

[203] Milcent A-S, Kadri A and Richir S 2022 Using Facial Expressiveness of a Virtual Agent to Induce Empathy in Users *International Journal of Human–Computer Interaction* **38** 240–52

[204] Pelachaud C 2009 Modelling multimodal expression of emotion in a virtual agent *Phil. Trans. R. Soc. B* **364** 3539–48

[205] Shiban Y, Schelhorn I, Jobst V, Hörnlein A, Puppe F, Pauli P and Mühlberger A 2015 The appearance effect: Influences of virtual agent features on performance and motivation *Computers in Human Behavior* **49** 5–11

[206] Chang Y-H, Maheswaran R, Levinboim T and Rajan V 2011 Learning and Evaluating Human-Like NPC Behaviors in Dynamic Games *Proceedings of the AAAI Conference on Artificial Intelligence and Interactive Digital Entertainment* **7** 8–13

[207] Garau M, Slater M, Pertaub D-P and Razzaque S 2005 The Responses of People to Virtual Humans in an Immersive Virtual Environment *Presence: Teleoperators & Virtual Environments* **14** 104–16

[208] Verhagen H, Johansson M and Eladhari M 2011 Model of Social Believable NPCs for Teacher Training *Proceedings of the 5th European Conference on Games Based Learning* European Conference on Games Based Learning (Athens, Greece)

[209] Iacovides I, Cox A, Kennedy R, Cairns P and Jennett C 2015 Removing the HUD: The Impact of Non-Diegetic Game Elements and Expertise on Player Involvement *Proceedings of the 2015 Annual Symposium on Computer-Human Interaction in Play* CHI PLAY '15 (New York, NY, USA: Association for Computing Machinery) pp 13–22

[210] Rosyid H A, Pangestu A Y and Akbar M I 2021 Can Diegetic User Interface Improves Immersion in Role-Playing Games? *2021 7th International Conference on Electrical, Electronics and Information Engineering (ICEEIE)* 2021 7th International Conference on Electrical, Electronics and Information Engineering (ICEEIE) (Malang, Indonesia: IEEE) pp 200–4

[211] Richter G, Raban D R and Rafaeli S 2015 Studying Gamification: The Effect of Rewards and Incentives on Motivation *Gamification in Education and Business* ed T Reiners and L C Wood (Cham: Springer International Publishing) pp 21–46

[212] Li L, Wang J, Xu G, Li M and Xie J 2015 The Study of Object-Oriented Motor Imagery Based on EEG Suppression *PLOS ONE* **10** e0144256

[213] Liang S, Choi K-S, Qin J, Pang W-M and Heng P-A 2014 Effective user training for motor imagery based brain computer interface with object-directed 3D visual display *2014 7th International Conference on Biomedical Engineering and Informatics* 2014 7th International Conference on Biomedical Engineering and Informatics pp 297–301

[214] Liang S, Choi K-S, Qin J, Pang W-M and Heng P-A 2016 Enhancing training performance for brain–computer interface with object-directed 3D visual guidance *Int J CARS* **11** 2129–37

[215] Gibson J J 1977 The theory of affordances *Perceiving, acting, and knowing: Toward an ecological psychology* ed R Shaw and J Bransford (Hillsdale, NJ: Erlbaum) pp 67–82

[216] Masoudi N, Fadel G M, Pagano C C and Elena M V 2019 A Review of Affordances and Affordance-Based Design to Address Usability *Proceedings of the Design Society: International Conference on Engineering Design* **1** 1353–62

[217] Schulz L, Ischebeck A, Wriessnegger S C, Steyrl D and Müller-Putz G R 2018 Action affordances and visuo-spatial complexity in motor imagery: An fMRI study *Brain and Cognition* **124** 37–46

[218] Leeuwis N, Paas A and Alimardani M 2021 Vividness of Visual Imagery and Personality Impact Motor-Imagery Brain Computer Interfaces *Front. Hum. Neurosci.* **15** 634748

[219] Mazarakis A and Bräuer P 2023 Gamification is Working, but Which One Exactly? Results from an Experiment with Four Game Design Elements *International Journal of Human–Computer Interaction* **39** 612–27

[220] Lee J and Lai K-Y 1991 What's in Design Rationale? *Human–Computer Interaction* **6** 251–80

[221] Lotte F, Jeunet C, Chavarriaga R, Bougrain L, Thompson D E, Scherer R, Mowla M R, Kübler A, Grosse-Wentrup M, Dijkstra K and Dayan N 2019 Turning negative into positives! Exploiting 'negative' results in Brain–Machine Interface (BMI) research *Brain-Computer Interfaces* **6** 178–89

[222] Mlinarić A, Horvat M and Šupak Smolčić V 2017 Dealing with the positive publication bias: Why you should really publish your negative results *Biochemia Medica* **27** 030201